\documentclass[prl,
reprint,
superscriptaddress,
amsmath,amssymb,
aps,twocols
]{revtex4-2}

\usepackage{graphicx}
\usepackage{dcolumn}
\usepackage{bm}
\usepackage{hyperref}
\usepackage[mathlines]{lineno}
\usepackage{comment}
\usepackage{graphicx,epsfig} 
\usepackage{amsmath}
\usepackage{amsfonts}
\usepackage{cancel}
\usepackage{bm}
\usepackage{amssymb}
\usepackage{amsmath,lipsum}
\usepackage{orcidlink}
\usepackage[normalem]{ulem}

\begin{document}
	
	
	\title{Phonon-driven nodal surface superconductivity of Fermi arcs}
	
	\author{Francesco Buccheri}
	\email{f.buccheri@ifw-dresden.de}
	\affiliation{Leibniz Institute for Solid State and Materials Research,\\
		IFW Dresden, Helmholtzstrasse 20, 01069 Dresden, Germany }
	\author{Alessandro De Martino}
	\affiliation{Department of Mathematics, City St George’s, University of London, London EC1V 0HB, United Kingdom}
	\author{Jeroen van den Brink}
	\email{j.van.den.brink@ifw-dresden.de}
	\affiliation{Leibniz Institute for Solid State and Materials Research,\\
		IFW Dresden, Helmholtzstrasse 20, 01069 Dresden, Germany }
	\affiliation{W\"urzburg-Dresden Cluster of Excellence ctd.qmat}
	
	\date{\today}
	
	\begin{abstract}
		According to recent observations, the topological surface states of Weyl semimetals may develop a superconducting gap, while bulk superconductivity remains absent. What drives the formation of this novel superconducting state is an open question. Here, we show that this phenomenon can arise from the interaction of Fermi arc electrons with both surface and bulk phonons in time-reversal-invariant Weyl semimetals. We identify two competing pairing channels, intra-arc and inter-arc, whose relative strength is governed by the efficiency of Coulomb screening at the surface. The combined effect of the Fermi arcs being disconnected and the weak screening of the Coulomb repulsion at the system's surface causes nodes to appear in the superconducting gap, as observed recently by photoelectron spectroscopy experiments on PtBi$_2$. This suggests manipulation of the Coulomb screening, e.g. by a surface layer coating,  as a pathway to engineer the critical temperature, as well as size and symmetry of the surface superconducting gap.  
	\end{abstract}
	
	\maketitle
	
	

	Recently a fascinating novel state of quantum matter was discovered in the Weyl semimetal PtBi$_2$: the topologically protected surface states of this Van der Waals material, consisting of electrons living on Fermi arcs, become superconducting, while the bulk crystal remains in its normal, non-superconducting state \cite{Veyrat2023,Kuibarov2024,Moreno2025}. Thus a topological surface superconductor (TSSC) materializes in PtBi$_2$, with angle resolved photoemission (ARPES)  measurements finding a superconducting gap opening up at the Fermi arcs below T$_c$$\sim$14 K.
	%
	%
	Also scanning tunneling experiments evidence robust superconducting
	gaps at the surface that are consistent with the superconductivity
	being carried by the Fermi arcs \cite{Schimmel2024,Moreno2025}. Strikingly,
	very recent ARPES measurements indicate the presence of nodes in the superconducting gap. The nodes belong to a manifold of Majorana modes located at the center of the Fermi arcs, and point toward PtBi$_2$ being an unconventional TSSC with $i$-wave pairing symmetry \cite{Changdar2025}. 
	Together with the unusual robustness of the surface superconducting
	state and its vortex phenomenology, these results place $\mbox{Pt}\mbox{Bi}_{2}$
	at the center of current efforts to realize and control intrinsic
	topological superconductivity.
	
	From the electronic structure perspective, $\mbox{Pt}\mbox{Bi}_{2}$
	is a layered, non-centrosymmetric semimetal hosting multiple Weyl
	points, pairwise connected by surface Fermi arcs. The detailed 
	electronic structure has been established in material-specific modeling
	and characterized experimentally \cite{Shipunov2020}. When the chemical
	potential lies near the Weyl nodes, the bulk density
	of states is strongly suppressed, while the surface density of states
	remains sizable, suggesting that a superconducting instability may indeed
	occur predominantly on the surface \cite{Vocaturo2024,Hoffmann2025,Zhang2025}.
	This scenario has motivated theoretical work ranging from surface-only Cooper
	instabilities in phenomenological arc models to mean-field studies
	of time-reversal-invariant Weyl semimetals \cite{Nomani2023}, as
	well as symmetry-based Ginzburg-Landau classifications tailored
	to $\mbox{Pt}\mbox{Bi}_{2}$ \cite{Bai2025,Waje2025} and variational
	approaches aimed at understanding the modulation of the order 
	parameter \cite{Trama2025}.
	
	A central open question is why the superconducting gap should be nodal
	on the Fermi arcs, despite the fact that a fully gapped state normally
	maximizes the condensation energy in BCS superconductors. Recent
	work \cite{Maeland2025,Maeland2025m} has provided numerical evidence
	for a microscopic mechanism: anisotropic electron-phonon coupling on
	Weyl-semimetal surfaces combined with statically screened Coulomb
	repulsion can stabilize nodal pairing on the arcs of this material,
	implying that Coulomb repulsion is not a small quantitative correction.
	
	Here we approach the problem from a complementary standpoint
	and show that the Coulomb-driven formation of nodes is a universal
	consequence of pairing on disconnected Fermi-arc segments, rather
	than a peculiarity of the microscopic lattice structure of $\mbox{Pt}\mbox{Bi}_{2}$.
	Building on a continuum description of time-reversal-invariant Weyl
	semimetals, we derive a surface pairing theory in which the effective
	interaction is generated by coupling of surface electrons to surface
	and bulk phonons, while lattice-specific characteristics (symmetries,
	mode content) enter only parametrically, in the same spirit as continuum
	surface theories used for topological surface states \cite{Armitage2018,Wan2011,Gorbar2015}.
	In our framework, multiple disconnected arcs are present and the arc
	index plays the role of an internal degree of freedom. This feature
	of the Fermi surface opens up the possibility of two different channels,
	associated with the two types of allowed momentum-conserving scattering
	processes. We find that intra-arc processes naturally generate conventional
	nodeless pairing, whereas inter-arc processes support
	momentum-odd pairing. When the Coulomb repulsion is weakly screened,
	it selectively suppresses small-momentum scattering, penalizing nodeless solutions and leaving the momentum-odd channel, involving large momentum transfer, as the only robust pairing instability.
	This produces nodes in the middle of the arc, in agreement with experimental observations \cite{Changdar2025}.
	As Fermi-arc states are most strongly localized near the center of
	the arc, the Coulomb repulsion is strongest precisely where a nodeless
	gap would otherwise be maximal, suppressing small-momentum-transfer
	pairing channels and favoring sign-changing, higher-angular-momentum
	solutions. 
	
	To quantitatively explore these concepts, we introduce a continuum description
	of surface Fermi-arc electrons coupled to bulk phonons and long-range
	Coulomb interactions. 
	We will consider the chemical potential as sufficiently close to the bulk nodes, so that the DOS of bulk electrons is negligible with respect to the surface DOS, which remains sizable also at charge neutrality \cite{Nomani2023}. This approach allows us to isolate the minimal
	physical ingredients responsible for pairing and node formation.
	We derive the phonon-mediated pairing kernel and the Coulomb form factor directly from the spatial profile of the surface states, and analyze how the penetration of surface states in the bulk reshapes the superconducting gap along the arc.
	
	\textit{Electronic continuum Hamiltonian.}
	Our starting point is a low-energy Hamiltonian of electrons in a Weyl semimetal. 
	Starting from the required discrete symmetries, we write a minimal two-band Hamiltonian describing the bulk electrons around the $\Gamma$ point
	\begin{equation}\label{eq:HWeyl}
		H\left(\boldsymbol{k}\right) =  D\left(k_{x}^{2}-k_{W}^{2}\right)\sigma^{x}+D\left(k_{y}^{2}-k_{W}^{2}\right)\sigma^{y}+\hbar v_{z} k_{z}\sigma^{z}. 
	\end{equation}
	where $\boldsymbol{k}=(k_x,k_y)$. Here the Pauli matrices act on an orbital degree of freedom which transforms as a scalar under the point group $C_{2z}$. The Hamiltonian \eqref{eq:HWeyl} is invariant under time-reversal, represented by the combination of the unitary matrix $\sigma^{x}$ and complex conjugation $\mathcal{K}$. Conversely, inversion is broken independently from the representation. The spectrum features two pairs of Weyl nodes, at momenta $(\pm k_W,\pm k_W,0)$. In Supplemental Material (SM) \cite{supp} Sec. III, we connect it to a tight-binding model defined on the cubic lattice, which is also instrumental in deriving the electron-phonon interaction Hamiltonian. For reference, we can estimate the parameters based on $\mbox{Pt}\mbox{Bi}_2$ as $k_W\approx 0.34 \mbox{\AA}^{-1}$, $Dk_W^2\approx\hbar v_{F}k_{W}/2\approx 0.38$eV \cite{Vocaturo2024}.
	
	We consider a boundary at $z=0$, so that the WS extends for $z<0$ and the vacuum for $z>0$. The most general boundary condition can be written by requiring that the matrix elements of the $z$-component of the current vanish on the boundary \cite{Witten2016,DeMartino2021}. As a consequence, our theory retains a free parameter $\alpha\in ]0,\pi/2]$, which controls the shape of the isoenergetic contours in the surface Brillouin zone and the localization of surface states. We find a two-dimensional band of surface states with wave function
	$\psi_{FA}=\sqrt{2\kappa}e^{\kappa z} \chi_-$, 
	where 
	\begin{equation}\label{eq:kappa}
		\kappa\left(\boldsymbol{k}\right)=\frac{D}{v_{z}\hbar}\left[\left(k_{W}^{2}-k_{x}^{2}\right)\sin\alpha-\left(k_{W}^{2}-k_{y}^{2}\right)\cos\alpha\right]
	\end{equation}
	is a momentum-dependent inverse penetration length and the spinor $\chi_-^T=\left(- e^{-i\alpha/2},e^{i\alpha/2}\right)/\sqrt{2}$ is fixed by the boundary conditions. This wavefunction identifies a pair of topological surface bands, with dispersion relation 
		\begin{equation}\label{eq:FAdispersion}
			\varepsilon\left(\boldsymbol{k}\right)= D\left[\left(k_{W}^{2}-k_{x}^{2}\right)\cos\alpha+\left(k_{W}^{2}-k_{y}^{2}\right)\sin\alpha\right]\,,
		\end{equation}
		protected by the bulk nodes.
	\begin{figure}
		\centering
		\centering{}\includegraphics[totalheight=0.2\textheight]{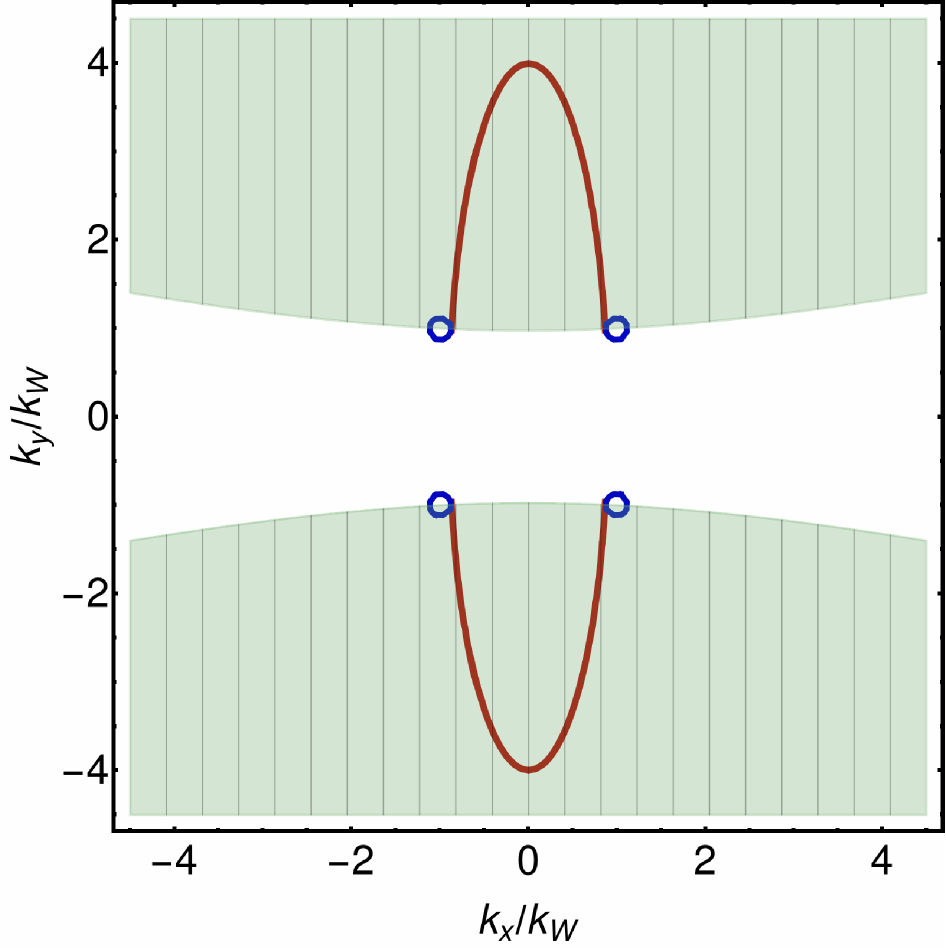}\caption{
			Fermi arcs, with fixed energy $E/\left(Dk_{W}^{2}\right)=0.25$, boundary parameter $\alpha=0.05$.
			The blue 
			circular contours delimit the projections of
			the bulk Fermi surface around each node. The shaded region indicates the region in which the surface states are normalizable for $z<0$.}\label{fig:Fermi arcs}
	\end{figure}
	The fermionic operator annihilating an electron in a surface state with momentum $\boldsymbol{k}$ is denoted by $c_{\boldsymbol{k}}$ in the following.

	\textit{Bulk and surface phonons}. Acoustic phonon modes are expected to be the most relevant at low temperatures, so we focus on these in our low-energy description and derive the mediated attractive electron-electron interaction. In order to obtain the universal features without sacrificing the essential physics, we consider an effective description of lattice vibrations as small deformations of a continuous elastic medium \cite{Landau1986}. In presence of a free surface, phonon modes include two linear combinations of the principal and secondary bulk waves, mixed by the boundary, a pure transverse shear mode and a surface-localized Rayleigh mode.
	
	The electron-phonon interaction can be derived by considering the variation of the hopping coefficients under a lattice strain \cite{Mahan}, see SM \cite{supp} Sec. III.C for details. It consists of a sum over contributions from the mentioned phonon modes $\lambda$ and over the allowed frequencies. We find
	\begin{eqnarray}\label{eq:Hep}
		H_{\text{ep}}&=&\sum_{\lambda}\intop\frac{d\boldsymbol{k}}{\left(2\pi\right)^{2}}\intop\frac{d\boldsymbol{q}}{\left(2\pi\right)^{2}}\sum_{\omega}\Bigg[g_{\lambda,\boldsymbol{k},\boldsymbol{q},\omega}a_{\omega,\boldsymbol{q}}^{(\lambda)}\left(t\right)
		\nonumber\\
		&&\qquad\quad  +g_{\lambda,\boldsymbol{k},\boldsymbol{q},\omega}^{*}a_{\omega,-\boldsymbol{q}}^{(\lambda)\dagger}\left(t\right)\Bigg]c_{\boldsymbol{k}+\boldsymbol{q}}^{\dagger}c_{\boldsymbol{k}}+\text{H.c.} .
	\end{eqnarray}
	in which the integration domain is restricted to the Fermi arcs, i.e., the momentum space region in which normalizable surface states exist, see \eqref{eq:kappa} and Fig. \ref{fig:Fermi arcs}.
	The form factors $g_{\lambda}$ result from the overlap between initial and final electron states and the envelop of the exchanged phonon. They are provided as intermediate expressions in SM \cite{supp} Sec. III.C, and are proportional to the square of the bare electron-phonon interaction strength $U_{\text{ep}}$, a free parameter in our model.
	As surface electrons and Rayleigh phonons are both localized around the surface, these form factors carry the largest weight, while processes involving the exchange of a bulk phonon are generally suppressed by a factor $1/\sqrt{L}$, with $L$ being the transverse size. Based on this observation only, one would expect that surface phonons are responsible for the buildup of superconductivity. However, we find that this is not the case, as explained below. 
	
	
	\textit{e-e interaction. } In order to proceed, we derive the effective electron-electron interaction and formulate a theory restricted to the Cooper channel, describing the scattering of electrons at time-reversed momenta. SM \cite{supp} Sec. III provides details about the reduction. The effective electron-electron interaction is written as
	\begin{equation}\label{eq:HeeCc}
		H_{ee}=
		\intop\frac{d\boldsymbol{k}}{\left(2\pi\right)^{2}}\intop\frac{d\boldsymbol{q}}{\left(2\pi\right)^{2}}V\left(\boldsymbol{k},\boldsymbol{q}\right)c_{\boldsymbol{k}+\boldsymbol{q}}^{\dagger}c_{-\boldsymbol{k}-\boldsymbol{q}}^{\dagger}c_{-\boldsymbol{k}}c_{\boldsymbol{k}}\,,
	\end{equation}
	in which the interaction potential $V=V_C+V_p$ results from the superposition of Coulomb repulsion and phonon-mediated attraction. The latter is written as 
	\mbox{$	V_p(\boldsymbol{k},\boldsymbol{q})=\sum_{\omega}\sum_{\lambda}V_{\lambda}\left(\boldsymbol{k},\boldsymbol{q},\omega\right)
		$,}
	with a sum over the phonon family $\lambda$ and the bands $\omega$. The potential $V_{\lambda}$ contains the overlap of initial and final states, and is modulated by the corresponding penetration depths, as well as by the spatial profile of the vibrational mode $\lambda$. The somewhat cumbersome expressions can be found in Sec. IV of SM \cite{supp}.
	Importantly, the number of bulk bands grows with the transverse size $L$, which compensates the $1/L$ suppression in the squared matrix elements, yielding a finite result. In fact, the larger phase space associated with bulk phonons yields the main contribution to the effective electron-electron potential for parameters modeled on $\mbox{Pt}\mbox{Bi}_2$, see Sec. II of SM \cite{supp} for details.
	This is the first ingredient of our theory: bulk lattice vibrations provide the "glue" in the electron-electron interaction. Despite the fact that electrons live on the surface, this aspect suggests that the Fermi arc superconductivity is not a purely two-dimensional phenomenon.
	
	As second ingredient, we account for the electron-electron interaction via the Coulomb term
	\begin{equation}\label{eq:VC}
		V_C(\boldsymbol{k},\boldsymbol{q})=\frac{4\pi C}{q+q_s}F_z\left(\boldsymbol{k},\boldsymbol{q}\right) 
	\end{equation}
	where $C=e^{2}/(4\pi\varepsilon_{0})$. Eq. \eqref{eq:VC} has the form of a Coulomb potential in two dimensions, modulated by a form factor $F_z$ which accounts for the varying penetration depth of the surface states, see SM \cite{supp} Sec. III.B for details.
	Static screening effects are generated by the surface carriers, from deviations of the bulk from charge neutrality or other bands not explicitly accounted for in the model, as well as from disorder, electron/hole pockets in the bulk 
	\cite{Skinner2014,DasSarma2015}, and environmental effects. As they are hard to compute exactly, we model them in \eqref{eq:VC} by a phenomenological Thomas-Fermi wavevector $q_s$, capturing the combined response of all low-energy screening channels \cite{Stern1967,DasSarma2015,Ghosh2020}.
	
\begin{figure}
		\centering
		\centering{}\includegraphics[totalheight=0.17\textheight]{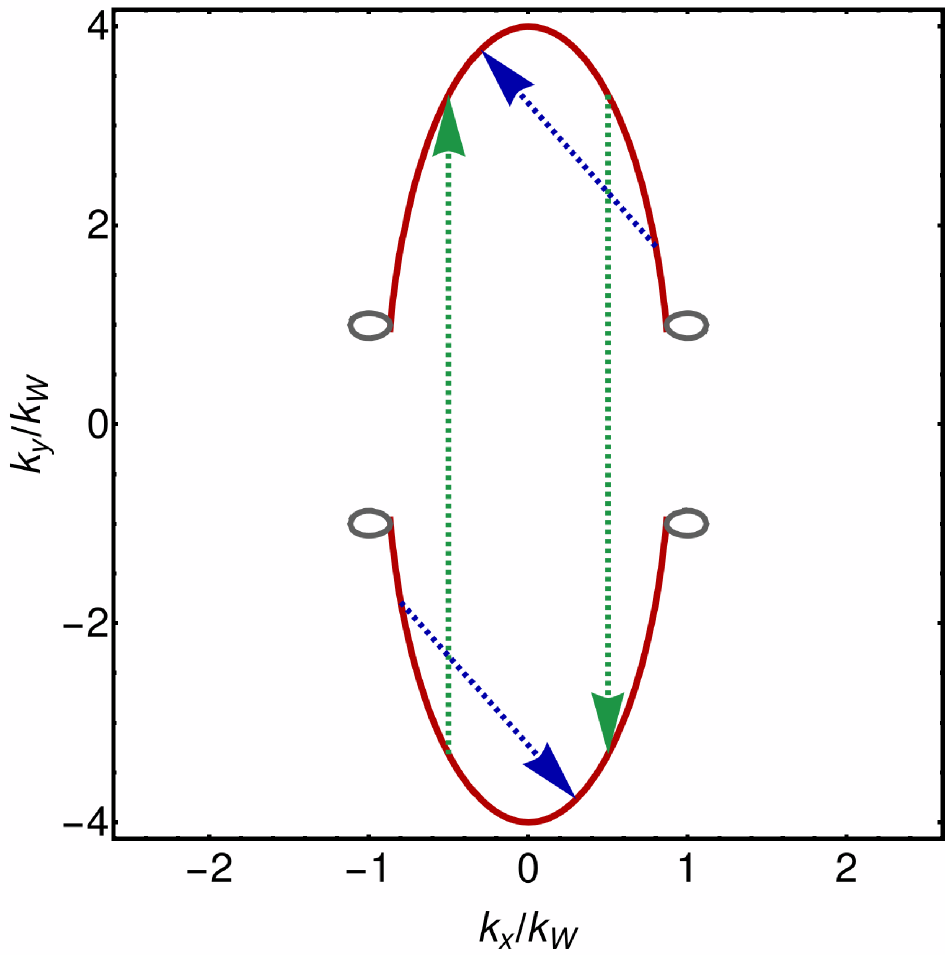}
		\hspace{0.15cm}\includegraphics[totalheight=0.17\textheight]{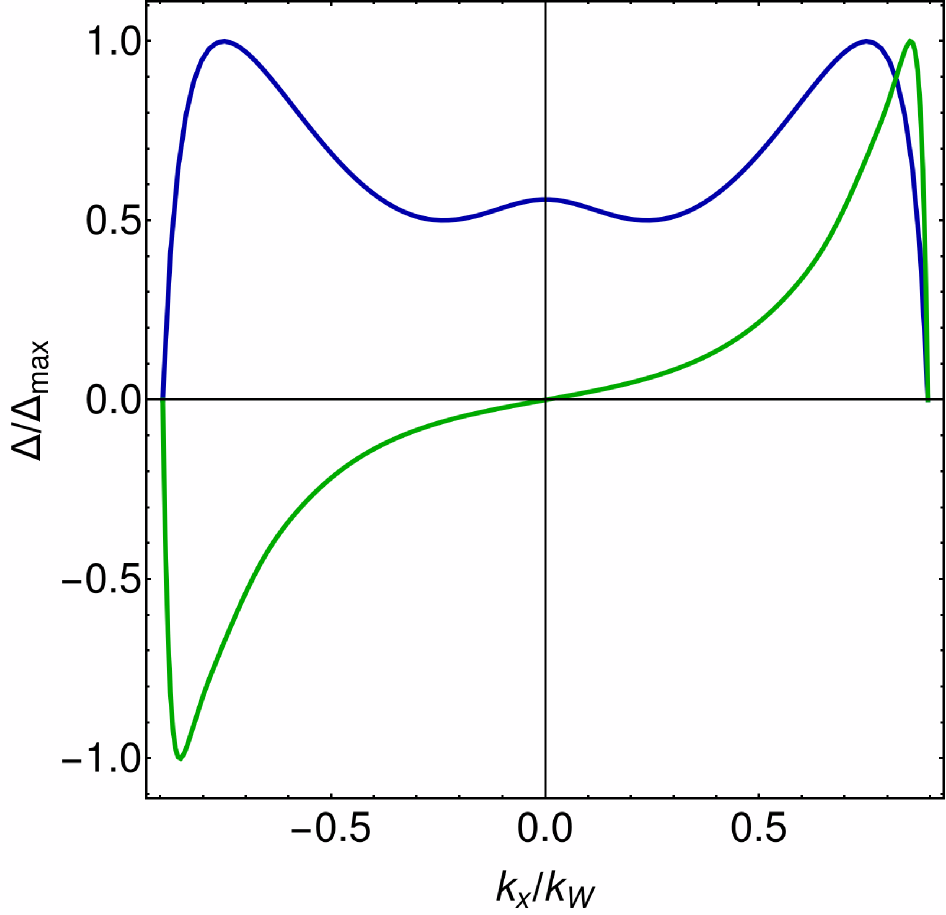}\caption{Left: illustration of the  intra- (blue) and inter-arc (green) scattering processes in the Cooper channel, with initial and final states on the Fermi arcs. Right: the nodal and node-less solutions (normalized to their maximum value) of the gap equation \eqref{eq:gapfull} projected to the Fermi surface, in presence of intra- (blue) or inter- (green) arc scattering. Here the boundary parameter $\alpha=0.05$ and the ep interaction strength $U_{\text{ep}}=0.025Dk_W^2$ have been used.}\label{fig:intervsintra}
	\end{figure}

	
	\textit{Surface superconductivity.  }
	A fixed-energy cut of the surface band \eqref{eq:FAdispersion} identifies two disconnected Fermi arcs \cite{Armitage2018}, represented in Fig. \ref{fig:Fermi arcs}. This illustrates the third ingredient of our model: a disconnected surface Fermi surface.
	To describe the associated physics, it is convenient to label the surface states by their energy $\varepsilon$, an index $\tau=\pm1$, identifying the top/bottom arc, and the momentum $k_x$. The set of quantum numbers $\left(\varepsilon,\tau,k_x\right)$ will be used in the following instead of $\boldsymbol{k}$, e.g., $c_{\varepsilon,\tau,k_x}$ annihilates a surface electron at energy $\varepsilon$ on the arc $\tau$ and momentum $k_x$.
	Momentum-conserving scattering processes around the Fermi energy are of two types, depicted in Fig. \ref{fig:intervsintra}, left panel: either the electrons switch arc (inter-arc) or stay on the same arc (intra-arc).
	Using the Nambu spinor
	$
	\Phi_{\varepsilon,k_x}=\left(c_{\varepsilon,+,k_{x}},-c_{\varepsilon,-,-k_{x}}^{\dagger}\right)^T
	$
	the Hamiltonian can be written in BdG form
	\begin{equation}
		\mathcal{H}=\intop d\varepsilon\intop\frac{dk_{x}}{2\pi}\,n(\varepsilon,k_x)\,
		\Phi^\dagger_{\varepsilon,k_x} \left(\begin{array}{cc}
			\xi_{\varepsilon} & \Delta_{\varepsilon,k_x}\\
			\Delta_{\varepsilon,k_x}^{*} & -\xi_{\varepsilon}
		\end{array}\right)\Phi_{\varepsilon,k_x}\label{eq:HBdgO},
	\end{equation}
	in which $\xi_\varepsilon=\varepsilon-\mu$ and the integration along $k_x$ is restricted to wavevectors within the arc. The function $\Delta$ is a superconducting gap generated by the effective interaction in the Cooper channel. The density of states (DOS) along either arc $n\left(\varepsilon,k_x\right)$ explicitly appears, see SM \cite{supp} Secs. III and V for further details. We can diagonalize the BdG Hamiltonian \eqref{eq:HBdgO} in terms of Bogoliubov quasiparticles \cite{Tinkham,Timm2025}, whose excitation spectrum $E=\sqrt{\xi_\varepsilon^2+\Delta^2}$ develops a gap when the integral equation
	\begin{align}\label{eq:gapfull}
		&	\Delta\left(\varepsilon;k_{x}\right)=\intop d\varepsilon'\intop\frac{dk'_{x}}{2\pi}\,n\,\left(\varepsilon',k'_{x}\right)\Bigg\{
		\\
		&\;-\mathcal{V}_{+,+}\left(\varepsilon,k_{x};\varepsilon',k'_{x}\right)\frac{\Delta\left(\varepsilon',k'_{x}\right)}{E\left(\varepsilon',k'_{x}\right)}\tanh\frac{\beta E\left(\varepsilon',k'_{x}\right)}{2}
		\nonumber\\
		&\;+\mathcal{V}_{+,-}\left(\varepsilon,k_{x};\varepsilon',k'_{x}\right)\frac{\Delta\left(\varepsilon',-k'_{x}\right)}{E\left(\varepsilon',-k'_{x}\right)}\tanh\frac{\beta E\left(\varepsilon',-k'_{x}\right)}{2}\Bigg\} \nonumber
	\end{align}
	has nontrivial solutions. The interaction potentials in the Cooper channel $\mathcal{V}$ in \eqref{eq:gapfull} and $V$ in \eqref{eq:HeeCc} are the same quantity, but expressed in terms of two different sets of variables. Specifically, the arguments of $\mathcal{V}$ are an initial state $(\varepsilon,\tau=+,k_x)$ and a final state $(\varepsilon',\tau',k'_x)$. The notation allow to differentiate between intra-arc $\mathcal{V}_{+,+}$ and inter-arc $\mathcal{V}_{+,-}$ interaction potentials. Importantly, the two terms enter the gap equation with a different sign. The interaction potentials are not constant in magnitude and in sign along the arcs and result from the sum of two terms: the phonon-mediated attraction provides a negative contribution, most pronounced when initial and final states are close to the end of the arc; the Coulomb repulsion generates a positive contribution, which is large for small momentum exchange. Only the former is necessary for establishing superconducting correlations in this model, while the latter selectively suppresses them.
	For the numerical solution in Fig. \eqref{fig:intervsintra}, we project the gap equation onto the Fermi arcs and iteratively solve it.
	As point inversion symmetry is broken in the bulk, pairing order parameters with different parity eigenvalues can mix. Nevertheless, the kernel of the projected gap equation commutes with a projected surface parity operation $\mathcal{P}$ implementing the map $k_x\to-k_x$. We can then classify the independent solutions according to the corresponding eigenvalue $\varpi=\pm1$, where $\Delta\left(-k_x\right)=\varpi \Delta\left(k_x\right)$. When the interaction is attractive everywhere along the arc, the sign structure in \eqref{eq:gapfull} implies that intra- and inter- arc channels favor gap solutions with opposite eigenvalues.
	If we only include the intra-arc scattering term $\mathcal{V}_{++}$, the gap equation \eqref{eq:gapfull} has the sign structure of the familiar BCS gap equation. For an attractive interaction, it is known that nodeless solutions have the largest condensation energy and transition temperature $T_c$, which we also verified in SM \cite{supp} Sec. V. Conversely, including inter-arc scattering $\mathcal{V}_{+-}$ only, we find that the $k_x$-odd self-consistent solution produces instead the lowest-energy state. 
	The numerical solutions for the intra- and inter-arc gaps are illustrated in Fig. \ref{fig:intervsintra}. 
	
\begin{figure}
		\includegraphics[height=0.2\textheight]{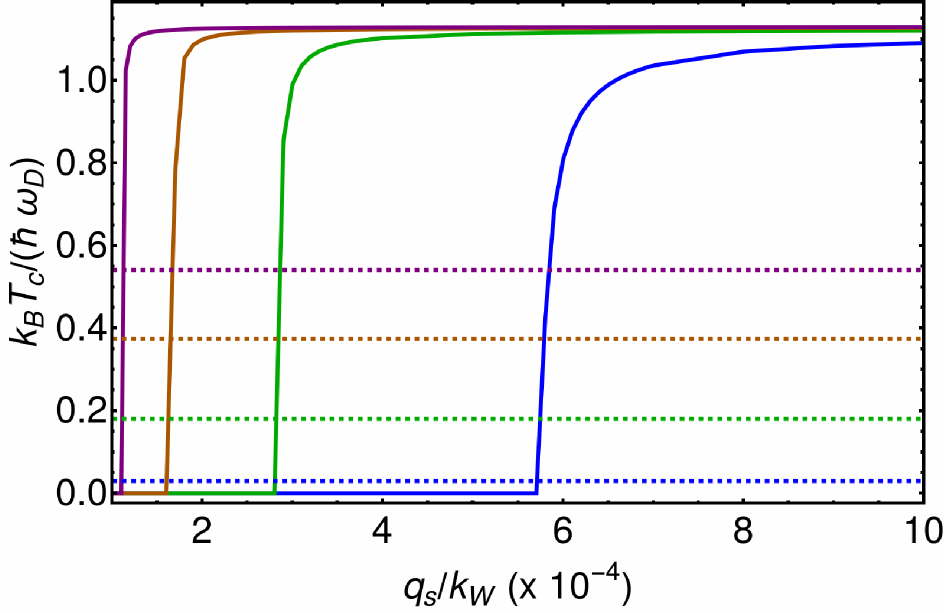}
		\caption{Critical temperature for intra- (solid) and inter- (dashed) arc pairing, as a function of the Thomas-Fermi screening momentum. For weak screening inter-arc pairing prevails and the gap is nodal. Here $\alpha=0.05$ and three values of the bare electron-phonon interaction rae $U_{\text{ep}}/Dk_W^2=0.025$ (blue), $0.035$ (green), $0.045$ (orange), $0.055$ (purple). \label{fig:Tc}}
	\end{figure}

	\textit{Critical temperatures and nodal superconductivity. } We now analyze the onset of superconductivity and how this is affected by the Coulomb interaction. To this end, we linearize the gap equations and again consider separately inter- and intra-arc channels. We obtain a homogeneous Fredholm integral equation of the second kind, which is solved numerically via discretization in momentum space, see also SM \cite{supp} Sec. V.B. The critical temperature is computed from the negative eigenvalue of largest magnitude $v_{min}^{(intra)}$ of the corresponding kernel
	\begin{equation}\label{Tintra}
		\frac{k_BT_{intra}}{\hbar\omega_{D}} 
		=\frac{2e^{\gamma}}{\pi}\exp\left\{ -\frac{1}{2\left|v_{min}^{(intra)}\right|}\right\} \;,
	\end{equation}
	in which $\omega_D$ is the Debye frequency and $\gamma$ is the Euler's constant. We observe that, decreasing the value of $q_s$, the minimal eigenvalue turns from negative to positive, the critical temperature goes to zero and the gap disappears, see Fig. \ref{fig:Tc}. 
	When only the inter-arc scattering is included, one obtains again a Fredholm equation with a different kernel. A useful approximation is to keep only the vertical transitions, with the exchanged momentum $\mathbf{q}=q_y\hat{y}$. Then the projected gap equation reduces to an algebraic relation, which only admits odd solutions. The gap remains odd when all inter-arc processes are included. The critical temperature is determined from the linearized gap equation in the same form \eqref{Tintra}, but with the negative eigenvalue of largest magnitude in the odd sector $v_{min}^{(inter)}$ appearing in the exponential, with details provided in Sec. V of \cite{supp}.
	The critical temperatures in the intra- and inter-arc channels can, in fact, be very different, because the former depends on the minimal eigenvalue of the intra-arc potential, hence, it is very sensitive to the Coulomb interaction and can be brought to zero by inefficient screening, e.g., for sufficiently low surface density of states. Conversely, the inter-arc potential is largely unaffected by the Coulomb repulsion, because only involves large-momentum transfer.
	The net result is visualized in Fig. \ref{fig:Tc}. For strong screening, the critical temperature for the intra-arc pairing is larger. Numerically, we observe that the gap is also generically larger, which in turn leads to a larger condensation energy for intra-arc pairing. Nevertheless, as soon as the Thomas-Fermi wavevector becomes small enough, the critical temperature of intra-arc pairing sharply drops.  
	In the weak screening regime, the odd inter-arc state is the only one surviving, as its critical temperature is very weakly sensitive to the Coulomb interaction.
	
	\textit{Discussion \& outlook. } Our analysis builds on the results of \cite{Nomani2023}, establishing that the surface can have a substantially higher superconducting temperature than the bulk of a Weyl semimetal. We focused therefore on the Fermi arc superconductivity, modeling the momentum dependence of the interactions from microscopic considerations. Our theory underlines the role of the bulk degrees of freedom in establishing the surface superconductivity, which can be nodeless or nodal, depending on the scattering processes at its origin. Nodal superconductivity appears as the generic outcome of pairing on disconnected surface Fermi surfaces, when both electron-phonon and competing weakly-screened long-range Coulomb interactions are present. Explicit inclusion of the Coulomb repulsion attributes therefore a subtly different role to the surface density of states: not only it establishes the onset of surface superconductivity, but it also screens the repulsive potential and favors a nodeless solution over a nodal one.
	
	Our results synthesize in a simplified setting the phenomenology observed $\mbox{Pt}\mbox{Bi}_{2}$: superconductivity on the surface is largely independent from the one in the bulk and is driven by Cooper pairing of Fermi-arc states, mediated predominantly by bulk phonons. As overlaps are strongly momentum dependent due to the spatial structure of the arcs, the effective interaction is maximal near the arc endpoints and suppressed around  the center of the arc, reflecting the varying penetration depth of  surface states into the bulk. As a result, the effective pairing interaction along the arc is intrinsically anisotropic and nonuniform \cite{Trama2025}.
	
	We solved the gap equation while retaining the momentum-dependence of the gap along the Fermi arc.
	Cooper pairs are built from electrons at time-reversed momenta in this model, but the disconnected Fermi surface allows to isolate two distinct pairing channels: intra-arc scattering preserves the ordering of arc indices between initial and final states and supports a nodeless, BCS singlet-like
		pairing state; inter-arc scattering exchanges the ordering of arc indices
		and supports a nodal order parameter, which follows from Fermi statistics upon this exchange. The corresponding solutions are triplet-like, with the arc index playing the role of an internal degree of freedom. 
	Coulomb repulsion selectively suppresses small momentum transfer,
	destabilizing nodeless gaps when screening is inefficient
	and leaving the nodal solution as the dominant instability.
	
	This provides a material-independent mechanism for the emergence
	of nodes in the superconducting gap on Fermi arcs, consistent with
	recent observations in $\mbox{Pt}\mbox{Bi}_{2}$. By isolating
	the minimal ingredients behind nodal pairing on Fermi arcs, our results
	provide a general principle for surface superconductivity in Weyl
	semimetals and a route toward engineering the nodal structure
	and topological interaction-driven states  \cite{Giwa2021,Changdar2025} via
	dielectric screening and interface design. Indeed, surface gating \cite{Potter2012,Park2017,Yang2025} 
	or a proximitized polarizable layer \cite{Fatemi2018,Steinke2020} modify the dielectric environment and can be used as active knobs to tune the value of $q_s$ in our model. This in turn provides a pathway to manipulate the Majorana hinge-states associated with the nodal $i$-wave superconductivity observed in PtBi$_2$ \cite{Changdar2025}. Note that the mechanism described here is very different from the proximity-induced surface superconductivity studied in \cite{Khanna2014,Chen2016,Chen2017}. 
	While our Hamiltonian is based on a type-I WS, the essential physics is rooted in the Fermi arcs. Therefore, our theory is also relevant for other Weyl systems, e.g., for interpreting the reports of surface superconductivity in the type-II WS $\mbox{Mo}\mbox{Te}_2$ \cite{Naidyuk2018,Li2018}. 
	
	\vspace{0.1cm}
	\textit{Acknowledgements. } We gratefully acknowledge stimulating exchanges with R. Fernandes, K. M\ae land, B. Trauzettel and R. Gonnelli. We acknowledge financial support by the Deutsche Forschungsgemeinschaft (DFG, German Research Foundation), through SFB 1143 Project A5 and the W{\"u}rzburg-Dresden Cluster of Excellence on Complexity and Topology in Quantum Matter-ct.qmat (EXC 2147, Project ID No. 390858490), and from the NQSTI (University of Catania, CUP E63C22002180006).
	
	
\newpage

\onecolumngrid

\pagebreak

\begin{center}
	{\bf Supplemental Material to "Phonon-driven nodal surface superconductivity of Fermi arcs"}
\end{center}

\setcounter{secnumdepth}{4}

\section{Lattice model}\label{app:lattice}
Starting from a tight-binding model, we derive here the continuum electronic Hamiltonian and the electron-phonon interaction terms used in the main text. Consider a cubic lattice with a total of $N$ sites and lattice constant $a$, on which we define the Hamiltonian
\begin{eqnarray}
	H & = & -\sum_{\boldsymbol{r}}\Bigg\{
	\frac{t}{2}\left(\psi_{\boldsymbol{r}}^{\dagger}\sigma^{x}\psi_{\boldsymbol{r}+\hat{x}}+
	\psi_{\boldsymbol{r}}^{\dagger}\sigma^{y}\psi_{\boldsymbol{r}+\hat{y}}\right)
	-\frac{t}{2}\psi_{\boldsymbol{r}}^{\dagger}\left(\sigma^{x}+\sigma^{y}-i\frac{t_{z}}{t}\sigma^{z}\right)\psi_{\boldsymbol{r}+\hat{z}}\label{eq:Hlattice}\\
	&  & \qquad\qquad\qquad\qquad-V_{0}\left[1+\cos\left(k_{W}a\right)\right]\psi_{\boldsymbol{r}}^{\dagger}\left(\sigma^{x}+\sigma^{y}\right)\psi_{\boldsymbol{r}}\Bigg\}+\mbox{H.c.}\nonumber 
\end{eqnarray}
where $t$ and $t_{z}$ are in- and out-of-plane hopping coefficients,
$V_{0}$ an onsite potential, which for simplicity is taken equal to the hopping
$t$ from now on and $k_{W}\in\left[0,\pi\right[$
is a parameter which determines the position of Weyl nodes. 
Expanding the electron operators as
\begin{equation}
	\psi_{\boldsymbol{r}}=\frac{1}{\sqrt{N}}\sum_{\boldsymbol{k}}e^{i\boldsymbol{k}\cdot\boldsymbol{r}/a}\psi_{\boldsymbol{k}}\,,
\end{equation}
we arrive at the Hamiltonian
\begin{eqnarray}
	H\left(\boldsymbol{k}\right) & = & t\left[1+\cos k_{W}-\cos k_{x}-\cos k_{z}\right]\sigma^{x}+t\left[1+\cos k_{W}-\cos k_{y}-\cos k_{z}\right]\sigma^{y}+t_{z}\sin k_{z}\sigma^{z} \;.\label{eq:Hlatt}
\end{eqnarray}
Its eigenvalues $E_{\pm}=\pm E\left(\boldsymbol{k}\right)$ are written as
\begin{eqnarray}
	E^{2}\left(\boldsymbol{k}\right) & = & t^{2}\left[1+\cos k_{W}-\cos k_{x}-\cos k_{z}\right]^{2}+t^{2}\left[1+\cos k_{W}-\cos k_{y}-\cos k_{z}\right]^{2}+t_{z}^{2}\sin^{2}k_{z}\;.
\end{eqnarray}
The four Weyl nodes are at $\left(\pm k_{W},\pm k_{W},0\right)$ with independent choices of $\pm$.

Defining the theory on a lattice is also instrumental to the derivation of an electron-phonon interaction Hamiltonian. This is achieved by modulating the dependence of the hopping amplitudes from the atomic positions. 
The hopping coefficient are taken, in first approximation, to be function of the distance
between the neighboring atoms. When the lattice site at position $\boldsymbol{r}$ is displaced by
$\mathbf{u}(\boldsymbol{r})$, the relative displacement between neighboring
sites modifies the hopping amplitudes.
As a consequence, the hopping coefficient along the $j$ direction is modified as
\begin{equation}
	t_{j}\to t_{j}+\delta t_{j}=t_{j}+U_{\text{ep}}\,\partial_{j}u_{j}(\mathbf{r})
\end{equation}
to leading order in the strain field \cite{Mahan}. Applying this procedure to the lattice
Hamiltonian (\ref{eq:Hlattice}), the electron-phonon interaction to linear order
is:
\begin{align}
	H_{\text{ep}}= & -\frac{U_{\text{ep}}}{2}\sum_{\mathbf{r}}\Big[\partial_{x}u_{x}(\mathbf{r})\,\psi_{\mathbf{r}}^{\dagger}\sigma_{x}\psi_{\mathbf{r}+\hat{x}}+\partial_{y}u_{y}(\mathbf{r})\,\psi_{\mathbf{r}}^{\dagger}\sigma_{y}\psi_{\mathbf{r}+\hat{y}}\nonumber \\
	& +\partial_{z}u_{z}(\mathbf{r})\,\psi_{\mathbf{r}}^{\dagger}(\sigma_{x}+\sigma_{y})\psi_{\mathbf{r}+\hat{z}}-i\,\partial_{z}u_{z}(\mathbf{r})\,\psi_{\mathbf{r}}^{\dagger}\sigma_{z}\psi_{\mathbf{r}+\hat{z}}\Big]+\text{H.c.}\label{eq:HepLattice}
\end{align}
Because of the spinor part of the surface states, see \eqref{eq:xipm}, the term proportional to $\sigma^{z}$ does not couple surface states, see \ref{app:epinteractions} for details about the contraction of the electron operators.

\section{Elastic continuum in presence of a surface}\label{app:elasticcontinuum}
In this Appendix, we review the theory of deformation of the elastic continuum in the presence of a free surface, i.e., not connected to a substrate. 
The dynamics of the displacement field $\boldsymbol{u}$ is described by the Lagrangian \cite{Bannov1994}
\begin{eqnarray}
	\mathcal{L} & = & \frac{1}{2}\intop dt\intop d\boldsymbol{r}\left[\rho\dot{\boldsymbol{u}}^{2}-\lambda u_{j,j}^{2}-2\mu u_{j,l}^{2}\right]\label{eq:Lph}
\end{eqnarray}
where $\rho$ is the volume mass density of the sample and $u_{j,l}$ is the linearized symmetric strain tensor
\begin{eqnarray}
	u_{j,l} & = & \frac{1}{2}\left(\frac{\partial u_{j}}{\partial r_{l}}+\frac{\partial u_{l}}{\partial r_{j}}\right)\;,\label{eq:strain}
\end{eqnarray}
For simplicity, here we have assumed isotropy of the elastic medium, by selecting the number of nonzero components of the stiffness tensor and writing them in terms of two Lamé coefficients $\lambda$, $\mu$ \cite{Landau1986}. These coefficients determine the two fundamental velocities of the phonons, the longitudinal $c_l=\sqrt{\left(\lambda+2\mu\right)/\rho}$ and the transverse $c_t=\sqrt{\mu/\rho}$, with $c_l>c_t$. 
By writing 
\begin{eqnarray}
	u_{j,l} & = & \left(u_{jl}-\frac{1}{3}\delta_{jl}\mbox{Tr}\left[u\right]\right)+\frac{1}{3}\delta_{jl}\mbox{Tr}\left[u\right]
\end{eqnarray}
we decompose the strain in a pure shear motion (traceless first term) and
a hydrostatic compression (second term). The equations of motion follow
from the stationarity of (\ref{eq:Lph}) and can be written in the
form
\begin{eqnarray}
	\ddot{\boldsymbol{u}} & = & c_{t}^{2}\nabla^{2}\boldsymbol{u}+\left(c_{l}^{2}-c_{t}^{2}\right)\nabla\left(\nabla\cdot\boldsymbol{u}\right) \;.\label{eq:ueom}
\end{eqnarray}
Within the linear elasticity theory, the stress tensor is written in terms of the strain
as 
\begin{eqnarray}\label{eq:stress}
	\sigma_{ik} & = & 
	\rho\left\{ \left(c_{l}^{2}-2c_{t}^{2}\right)\nabla\cdot\boldsymbol{u}\delta_{ik}+2c_{t}^{2}u_{ik}\right\} 
\end{eqnarray}

The free surface at $z=0$ is described by stress-free boundary conditions \cite{Landau1986,Bannov1994,Giraud2011,Buccheri2022}
\begin{eqnarray}
	&  & \sigma_{z,x}\left(z=0,\boldsymbol{r}\right)=\sigma_{z,y}\left(z=0,\boldsymbol{r}\right)=\sigma_{z,z}\left(z=0,\boldsymbol{r}\right)=0\label{eq:phononbc}.
\end{eqnarray}
The eigenmodes can be written as the solution of the eigenvalue problem defined by \eqref{eq:Lph} with the boundary conditions \eqref{eq:phononbc}. The resulting phonon spectrum is divided into sectors, according to frequency and conserved momentum parallel to the interface, denoted by $\boldsymbol{q}=\left(q_{x},q_{y}\right)$. At large frequencies $\omega>c_lq$ the modes have a bulk character. They are modeled by free waves reflected at the surface in a half-infinite geometry or by standing waves in a slab. Depending on their oscillation direction on the surface with respect to the wavector, we divide them into surface longitudinal $l$ (or dilatational), surface transverse $t$ (or flexural) and pure shear $s$ modes. At small frequencies $\omega<c_tq$, instead, the only allowed mode is a surface-localized Rayleigh mode, with a purely two-dimensional character. Finally, at intermediate frequencies $c_tq<\omega<c_lq$ the modes have a mixed character and can be written as a superposition of a reflected wave and an exponentially localized distortion. 

It is easiest to represent the normal modes in cylindrical coordinates, with the
$\boldsymbol{q}=\left(q_{x},q_{y}\right)$ direction representing
the wavevector along the surface $\boldsymbol{u}^{T}=\left(\begin{array}{ccccc}
	u_{k} & , & u_{\phi} & , & u_{z}\end{array}\right)$. For later reference, define the notation 
\begin{equation}\label{eq:omegaltqlt}
	\omega_{l,t}=\frac{\omega}{c_{l,t}}\qquad\qquad\qquad\qquad q_{l,t}=\sqrt{\frac{\omega^{2}}{c_{l,t}^{2}}-q^{2}}
\end{equation}
and $\nu=c_{t}/c_{l}$. We introduce a transverse size $L$ in the $z$
direction, which we use as a cutoff and that will be at later stage sent to infinity. The modes in the various frequency ranges are:
\begin{description}
	\item [{Large frequencies}] for $\omega_{l}^{2}>q^{2}$ we have three
	independent modes. The first is the surface ``longitudinal'' (or
	``dilatational'' in the slab) mode 
	\begin{eqnarray}
		\boldsymbol{w}_{\omega,\boldsymbol{q}}^{(l)}\left(z\right) & = & \sqrt{\frac{2}{L}}\frac{2q^{2}q_{t}}{\sqrt{\omega_{t}^{6}-4q^{2}q_{t}^{2}\left(\omega_{t}^{2}-\omega_{l}^{2}\right)}}\left(\begin{array}{c}
			\cos\left(q_{l}z\right)+\frac{q_{t}^{2}-q^{2}}{2q^{2}}\cos\left(q_{t}z\right)\\
			0\\
			i\left[\frac{q_{l}}{q}\sin\left(q_{l}z\right)+\frac{q^{2}-q_{t}^{2}}{2q_{t}q}\sin\left(q_{t}z\right)\right]
		\end{array}\right)\label{eq:longitudinal}
	\end{eqnarray}
	The name stems from the fact that on the surface $z=0$ there is only
	a modulation in the direction of the wavevector. The second mode is
	a surface ``transverse'' (or ``flexural'' in a slab) mode
	\begin{eqnarray}
		\boldsymbol{w}_{\omega,\boldsymbol{q}}^{(t)}\left(z\right) & = & \sqrt{\frac{2}{L}}\frac{2q^{2}q_{l}}{\sqrt{\omega_{t}^{4}\omega_{l}^{2}-4q^{4}\left(\omega_{t}^{2}-\omega_{l}^{2}\right)}}\left(\begin{array}{c}
			\frac{q^{2}-q_{t}^{2}}{2q_{l}q}\sin\left(q_{l}z\right)+\frac{q_{t}}{q}\sin\left(q_{t}z\right)\\
			0\\
			i\left[\frac{q_{t}^{2}-q^{2}}{2q^{2}}\cos\left(q_{l}z\right)+\cos\left(q_{t}z\right)\right]
		\end{array}\right)\label{eq:transverse}
	\end{eqnarray}
	On the surface, the displacement associated to this mode has only
	a component perpendicular to the wavevector 
	$\boldsymbol{q}\cdot\boldsymbol{w}_{\omega,q}^{(s)}\left(z=0\right)=0$
	and to the surface itself. These two modes are the two independent superpositions of principal and secondary bulk waves, which are mixed by the boundary \cite{Landau1986}. Finally, we have a pure ``shear'' mode
	for $\omega_{t}>q$
	\begin{eqnarray}
		\boldsymbol{w}_{\omega,\boldsymbol{q}}^{(s)}\left(z\right) & = & \sqrt{\frac{2}{L}}\left(\begin{array}{c}
			0\\
			\cos\left(q_{t}z\right)\\
			0
		\end{array}\right)\label{eq:shear}
	\end{eqnarray}
	This is still a transverse mode on the surface because
	$\boldsymbol{q}\cdot\boldsymbol{w}_{\omega,q}^{(s)}\left(z=0\right)=0$,
	but the displacement is now along the surface.
	\item [{Small frequencies}] for $\omega_{t}^{2}<q^{2}$ only the Rayleigh
	mode is present. It is exponentially localized around the surface
	and its dispersion is linear in the wavevector
	\begin{eqnarray}
		\omega_{R} & = & c_{R}q
	\end{eqnarray}
	where the Rayleigh phonon velocity is $c_{R}=\xi_{R}c_{t}$, with
	$\xi_{R}=\sqrt{\frac{8}{3}-\frac{4}{3}\sqrt{12\nu-2}\cos\left[\frac{1}{3}\arccos\frac{17-45\nu}{\left(12\nu-2\right)^{3/2}}\right]}$ \cite{Giraud2011}. 
	Using $\kappa_{l}=\sqrt{1-\xi_{R}^{2}\nu^{2}}q$ and $\kappa_{t}=\sqrt{1-\xi_{R}^{2}}q$,	the mode is written as
	\begin{eqnarray}
		\boldsymbol{w}_{\omega,\boldsymbol{q}}^{(R)}\left(z\right) & = & \mathcal{N}_{\omega,\boldsymbol{q}}^{(R)}\left(\begin{array}{c}
			\frac{2-\xi_{R}^{2}}{2\sqrt{1-\xi_{R}^{2}\nu^{2}}}e^{\sqrt{1-\xi_{R}^{2}\nu^{2}}qz}-\sqrt{1-\xi_{R}^{2}}e^{\sqrt{1-\xi_{R}^{2}}qz}\\
			0\\
			-i\frac{2-\xi_{R}^{2}}{2}e^{\sqrt{1-\xi_{R}^{2}\nu^{2}}qz}+ie^{\sqrt{1-\xi_{R}^{2}}qz}
		\end{array}\right)\label{eq:Rayleighx}
	\end{eqnarray}
	with the normalization
	\begin{eqnarray}
		\mathcal{N}_{\omega,\boldsymbol{q}}^{(R)} & = &
		\sqrt{\frac{2q\left(1-\xi_{R}^{2}\nu^{2}\right)\sqrt{1-\xi_{R}^{2}}}{\left(1-\xi_{R}^{2}\right)\left(2-\xi_{R}^{2}\nu^{2}\right)+\left[1-\xi_{R}^{2}\nu^{2}-2\sqrt{\left(1-\xi_{R}^{2}\nu^{2}\right)\left(1-\xi_{R}^{2}\right)}\right]\left(2-\xi_{R}^{2}\right)}}=\sqrt{2q}\mathcal{N}_{0}^{(R)}\label{eq:NRx}
	\end{eqnarray}
	This function vanishes as $q^{1/2}$ for $q\to0$.
	\item [{Intermediate frequencies}] in the regime $\omega_{l}^{2}<q^{2}<\omega_{t}^{2}$,
	we still have the ``s'' mode, but neither ``l'' nor ``t'' modes.
	Instead, we have a ``mixed'' mode
	\begin{eqnarray}
		\boldsymbol{w}_{\omega,\boldsymbol{q}}^{(m)}\left(z\right) & = & \frac{1}{\omega_{t}\sqrt{\frac{L}{2}\left[1+\frac{\left(q_{t}^{2}-q^{2}\right)^{4}}{16\kappa_{l}^{2}q_{t}^{2}q^{4}}\right]}}\left(\begin{array}{c}
			-\frac{q_{t}^{2}-q^{2}}{2\kappa_{l}}e^{\kappa_{l}z}+q_{t}\sin\left(q_{t}z\right)-\frac{\left(q_{t}^{2}-q^{2}\right)^{2}}{4q^{2}\kappa_{l}}\cos\left(q_{t}z\right)\\
			0\\
			i\left[\frac{q_{t}^{2}-q^{2}}{2q}e^{\kappa_{l}z}+q_{x}\cos\left(q_{t}z\right)+\frac{\left(q_{t}^{2}-q_{x}^{2}\right)^{2}}{4\kappa_{l}qq_{t}}\sin\left(q_{t}z\right)\right]
		\end{array}\right)\label{eq:mixed}
	\end{eqnarray}
	which is partially localized around the surface, but extends in the
	bulk.
\end{description}
It can be readily verified that these modes satisfy the equations of motion (\ref{eq:ueom}) and the
boundary conditions (\ref{eq:phononbc}). We employ throughout the paper the normalization
$
\intop_{-L}^0 dz\,\boldsymbol{w}_{\boldsymbol{q}}^{\dagger}\cdot\boldsymbol{w}_{\boldsymbol{q}} =  1
$
so the $\boldsymbol{w}$ modes have dimensions $\mbox{length}^{-1/2}$. In numerical evaluations we employ the values $c_{l}\approx2.5\times10^{3}\mbox{m}/\text{s}$ and $c_{t}\approx1.25\times10^{3}\mbox{m}/\text{s}$, approximately modeled on the values extracted from recent laser-pulse measurements \cite{Liang2026}.


In order to properly define the sums over $\omega$ appearing in the interaction, we consider a large, yet finite, transverse size $L$ along the $z$ direction. Therefore a sum over $\omega$ is simply a
sum over the transverse bands. 
In such a geometry, the "longitudinal" (or "compressional") modes are usually
referred to as "symmetric" modes, while the "transverse" (or "flexural") are
usually dubbed "antisymmetric" waves \cite{Auld1974}.
At finite $L$, each mode is constrained by a single transcendental
equation \cite{Bannov1994,Giraud2012,Auld1974}. In the thick slab limit, $L\to\infty$ at fixed $q$,
there are two sets of solutions of the quantization equation for each phonon family. For $l$ modes, either $q_{t,n}=\frac{2n\pi}{L}$ or $q_{l,n}=\frac{\left(2n+1\right)\pi}{L}$. For $t$ modes, either $q_{t,n}=\frac{\left(2n+1\right)\pi}{L}$ or $q_{l,n}=\frac{2n\pi}{L}$.
We can therefore trade a sum over modes for an integral over $q_{l}$ or $q_{t}$. The two are not independent because of \eqref{eq:omegaltqlt}, so $dq_{l,t}=\frac{\omega}{c_{l,t}^{2}q_{l,t}}d\omega$. 
As a result, the sum over frequencies is properly defined via
\begin{eqnarray}\label{eq:frequencysum}
	\sum_{\omega} & \equiv & \sum_{n}\,\approx\,\frac{L}{2\pi}\left(\intop dq_{t}+\intop dq_{l}\right)\,=L\intop^{\omega_{D}}\frac{d\omega}{2\pi}\,N\left(\omega\right)
\end{eqnarray}
where the density of phonon states at given frequency per unit length is
$
N\left(\omega\right) =  \omega\left(\frac{1}{c_{l}^{2}q_{l}}+\frac{1}{c_{t}^{2}q_{t}}\right).
$
The integral \eqref{eq:frequencysum} features a high-frequency cutoff, namely, a Debye frequency $\omega_{D}$. 
This is a phenomenological parameter in our model, which is estimated from the corresponding measured temperature $T_{D}=143$K \cite{Xing2019,Xing2020,Bashlakov2022}, from which 
$\omega_{D}\approx12.5$meV
.
The transverse momentum of $s$ modes is instead quantized as $k_t=\pi n/L$, $n=0,1,\ldots$ in the limit $L\to\infty$. One writes the sum in the same way, but with density of phonon states 
$N_s\left(\omega\right) =\frac{2\omega}{c_t^2k_t}$.
The factor $L$ in front of the integral in \eqref{eq:frequencysum} exactly cancels the corresponding
$1/L$ arising from the squared overlaps containing the bulk phonons \eqref{eq:longitudinal}, \eqref{eq:transverse}, \eqref{eq:shear}, producing a finite result.

After quantization, the displacement field is expanded into normal modes $\boldsymbol{w}_{\omega,\boldsymbol{q}}^{(\lambda)}$ as
\begin{equation}
	\boldsymbol{u}\left(\boldsymbol{r},z\right) =  \frac{1}{2}\sum_{\lambda,\omega}\sqrt{\frac{\hbar}{2\omega\rho}}\intop\frac{d\boldsymbol{q}}{\left(2\pi\right)^{2}}e^{i\boldsymbol{q}\cdot\boldsymbol{r}}\Big[a_{\omega,\boldsymbol{q}}^{(\lambda)}\left(t\right)\boldsymbol{w}_{\omega,\boldsymbol{q}}^{(\lambda)}\left(z\right) +a_{\omega,-\boldsymbol{q}}^{(\lambda)\dagger}\left(t\right)\boldsymbol{w}_{\omega,-\boldsymbol{q}}^{(\lambda)*}\left(z\right)\Big]\label{eq:ufield-2-intdk}
\end{equation}
with bosonic ladder operators $a_{\omega,\boldsymbol{q}}^{(\lambda)}$ satisfying canonical commutation relations. For compactness of notation, we understand that the mode index $\lambda=R,m,s,t,l$ is restricted according to the momentum and frequency regime, as described above. 
Finally, we write here for reference the phononic Hamiltonian
\begin{equation}
	H_{ph}=\intop\frac{d\boldsymbol{q}}{\left(2\pi\right)^{2}}\left\{ \hbar\omega_{R}\left(\boldsymbol{q}\right)\Theta\left(c_{t}q-\omega_{R}\left(\boldsymbol{q}\right)\right)a_{\boldsymbol{q}}^{(R)\dagger}a_{\boldsymbol{q}}^{(R)} +\sum_{c_{t}q<\omega<c_{l}q}\sum_{\lambda=m,s}\hbar\omega a_{\omega,\boldsymbol{q}}^{(\lambda)\dagger}a_{\omega,\boldsymbol{q}}^{(\lambda)}
	+\sum_{\omega>c_{l}q}\sum_{\lambda=l,t,s}\hbar\omega a_{\omega,\boldsymbol{q}}^{(\lambda)\dagger}a_{\omega,\boldsymbol{q}}^{(\lambda)}\right\} \;. \label{eq:Hph}
\end{equation}

\section{Boundaries and interactions in the electronic model}\label{app:electrons}
In this Appendix, we provide further details about the electronic model and the interaction terms in the Hamiltonian.

\subsection{Zero-current boundary condition and Fermi arcs}
The most general momentum-independent family of boundary conditions that ensures
hermiticity of the Hamiltonian (1) in the main text \cite{main}, when defined in
the half-space $z<0$, takes the form \cite{Witten2016,Burrello2019,Buccheri2022}
\begin{equation}\label{eq:BC}
	B\,\psi\left(\boldsymbol{k},z=0\right) = - \psi\left(\boldsymbol{k},z=0\right)\;.
\end{equation}
Here the boundary matrix depends on the real angular variable $\alpha$ and reads
\begin{equation}\label{eq:Balpha}
	B \;=\; \sigma^{x}\cos\alpha+\sigma^{y}\sin\alpha \;=\; \left(\begin{array}{cc}
		0 & e^{-i\alpha}\\
		e^{i\alpha} & 0
	\end{array}\right).
\end{equation}
	To see this, we have to use that translation invariance along $z$ is broken, so $k_{z}$ is no longer a good quantum number. Requiring
	$\left\langle \psi_{1}\big|H\psi_{2}\right\rangle
	=\left\langle H\psi_{1}\big|\psi_{2}\right\rangle$
	for every pair of states $\psi_{1,2}$ in the domain forces the boundary term
	$-i\hbar v_{z}\,\psi_{1}^{\dagger}(z=0)\sigma^{z}\psi_{2}(z=0)$ to vanish. This condition states that all matrix elements of the current component $j_{z}=ev_{z}\sigma^{z}$ perpendicular to the boundary vanish at $z=0$.
	The two eigenvalues of $\sigma^{z}$ correspond to the components of the spinor carrying flux toward and away from the surface, so this condition requires the two to enter with equal weight, i.e., that the surface reflects perfectly, without absorbing. The only freedom left is the phase acquired upon reflection, so the admissible states form the one-parameter family 
	\begin{equation} \label{eq:xipm}
		\chi_{\pm}  =  \frac{1}{\sqrt{2}}\left(\begin{array}{c}
			\pm e^{-i\alpha/2}\\
			e^{i\alpha/2}
		\end{array}\right)
	\end{equation}
	which are indeed the eigenstates of \eqref{eq:BC}. The $\pm$ sign there is a convention, as $B\to-B$ under $\alpha\to\alpha+\pi$. For $0<\alpha\le\pi/2$, this choice ensures normalizable eigenstates. We further restrict throughout this work to the momentum-independent case, so that $\alpha$ is a single global parameter. 
	Finally, with time reversal implemented as $\mathcal{T}=\sigma^{x}\mathcal{K}$,
	one readily sees that $\mathcal{T}B\mathcal{T}^{-1}=B$ for every
	$\alpha$.

\begin{figure}
	\centering{}
	\includegraphics[totalheight=0.19\textheight]{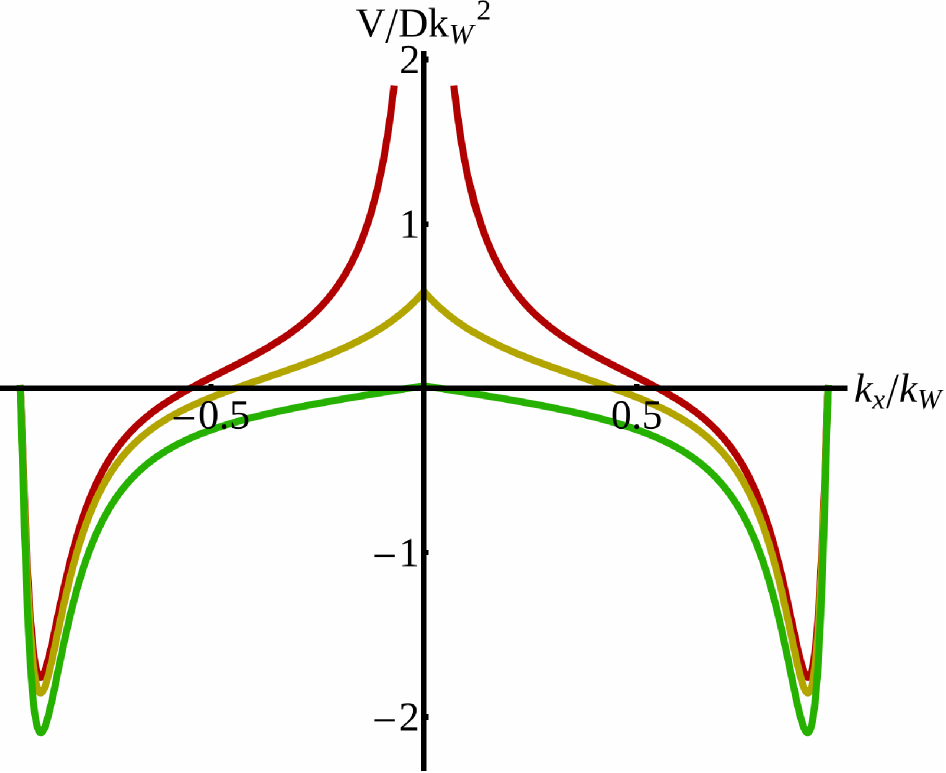}
	\hspace{0.5cm}
	\includegraphics[totalheight=0.19\textheight]{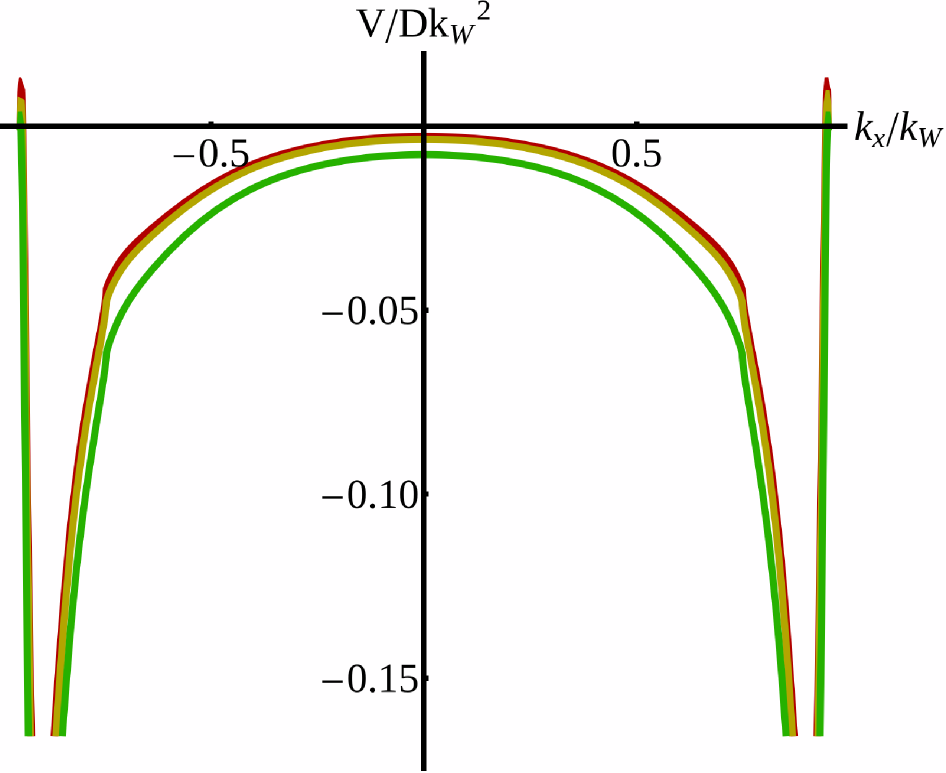}
	\caption{Left: inter-arc $k_x\to - k_x$ scattering potential. Left: inter-arc $\tau\to-\tau$ scattering potential, for various values of the Thomas-Fermi wavevector, $q_s/k_W=5$ (green), $0.5$ (yellow), $0.025$ (red). \label{fig:FAopen}}
\end{figure}

At fixed energy, we retain $k_x$ as an independent label and write $k_y$ from the dispersion relation of the Fermi arcs as
\begin{equation}\label{eq:ky}
	k_{y}\left(E,k_{x}\right)=k_{W}\sqrt{1+\left(1-\frac{k_{x}^{2}}{k_{W}^{2}}\right)\cot\alpha-\frac{E}{Dk_{W}^{2}\sin\alpha}}\qquad\qquad\left|k_{x}\right|<k_{end}
\end{equation}
As shown in Fig. \ref{fig:FAopen}, the Fermi surfaces around the nodes $(k_{x}=\pm k_{W},k_{y}=-k_{W})$ are connected by the arc $k_{y}=-k_{y}\left(\varepsilon,k_{x}\right)$ and the Fermi surfaces around the nodes $(k_{x}=\pm k_{W},k_{y}=k_{W})$ by the arc $k_{y}=+k_{y}\left(\varepsilon,k_{x}\right)$. The arcs merge with bulk states when $\kappa\left(E,k_{x}\right)=0$, i.e., when their penetration depth diverges. This takes place at the momenta $k_{x}=\pm k_{end}(\varepsilon)$, where
\begin{equation}\label{eq:kend}
	k_{end}\left(E\right)=k_{W}\sqrt{1-\frac{E\cos\alpha}{Dk_{W}^{2}}} \;.
\end{equation}
In this model, the ratio between the elongation along $y$ and the base of the arc is approximately \mbox{ $h=\frac{1}{2}\sqrt{1+\cot\alpha}+\mathcal{O}\left(\frac{\varepsilon}{Dk_{W}^{2}}\right)$}, for a chemical potential around the Weyl nodes. Importantly, the penetration depth is not constant along the arcs: it diverges at the arc ends and is maximal at the center, where \mbox{$\kappa\left(\varepsilon,k_{x}=0\right)=\frac{Dk_{W}^{2}-\varepsilon\cos\alpha}{\hbar v_{z}\sin\alpha}$}. The momentum-dependent DOS along the arc is
\begin{equation}\label{eq:n}
	n\left(\varepsilon,k_x\right) = \left[4\pi D k_y\left(\varepsilon,k_x\right) \sin\alpha \right]^{-1} \;,
\end{equation}
while the corresponding integrated DOS at given energy $\varepsilon$ is
\begin{equation}
	N\left(\varepsilon\right)=\intop\frac{d\boldsymbol{k}}{\left(2\pi\right)^{2}}\delta\left(\varepsilon-D\left(k_{W}^{2}-k_{x}^{2}\right)\cos\alpha-D\left(k_{W}^{2}-k_{y}^{2}\right)\sin\alpha\right)
	=\frac{1}{2\pi^{2}D}\frac{\arcsin \left( q_{\mbox{end}}\right)}{\sqrt{\sin\alpha\cos\alpha}}\;,
\end{equation}
where
\begin{equation}
	q_{\mbox{end}}=\sqrt{\frac{\cos\alpha\left(1-\frac{\varepsilon}{Dk_{W}^{2}}\cos\alpha\right)}{\sin\alpha+\cos\alpha-\frac{\varepsilon}{Dk_{W}^{2}}}} \;.
\end{equation}

\subsection{Coulomb interaction}\label{app:Coulomb}
Now we derive the bare Coulomb interaction acting on pairs of Fermi arc electrons, eq. (6) in the main text. The starting point is the textbook electron-electron interaction \cite{Mahan,Timm2025}
\begin{equation}
	H_{C}=\frac{1}{2}\intop d^{2}\boldsymbol{r}_{1}\intop dz_{1}\intop d^{2}\boldsymbol{r}_{2}\intop dz_{2}\psi^{\dagger}\left(\boldsymbol{r}_{1}\right)\psi^{\dagger}\left(\boldsymbol{r}_{2}\right)\mathcal{V}_{C}\left(\left|\boldsymbol{r}_{1}-\boldsymbol{r}_{2}\right|\right)\psi\left(\boldsymbol{r}_{2}\right)\psi\left(\boldsymbol{r}_{1}\right).
\end{equation}
The Coulomb pairwise interaction increases the energy by $V_{C}\left(x,y,z\right)=\frac{C}{\sqrt{r^{2}+z^{2}}}$, with the constant $C$ defined in the main text. Performing the Fourier transform along the surface direction and moving to center-of-mass and relative coordinates, we obtain
\begin{eqnarray}
	H_{C}&=&\frac{1}{2\mathcal{S}^{4}}\sum_{\boldsymbol{k}_{1}\ldots\boldsymbol{k}_{4}}\intop d^{2}\boldsymbol{R}\intop d^{2}\boldsymbol{r}\intop dZ\intop dze^{i\left(\boldsymbol{k}_{4}-\boldsymbol{k}_{1}+\boldsymbol{k}_{3}-\boldsymbol{k}_{2}\right)\cdot\boldsymbol{R}+i\left(\boldsymbol{k}_{4}-\boldsymbol{k}_{1}-\boldsymbol{k}_{3}+\boldsymbol{k}_{2}\right)\cdot\frac{\boldsymbol{r}}{2}}\\
	&&\qquad\qquad\times\psi_{\boldsymbol{k}_{1}}^{\dagger}\left(Z+\frac{z}{2}\right)\psi_{\boldsymbol{k}_{2}}^{\dagger}\left(Z-\frac{z}{2}\right)V_{C}\left(\boldsymbol{r},z\right)\psi_{\boldsymbol{k}_{3}}\left(Z-\frac{z}{2}\right)\psi_{\boldsymbol{k}_{4}}\left(Z+\frac{z}{2}\right)
	\nonumber
\end{eqnarray}
in which $\mathcal{S}$ is an area and $Z\le-{\left|z\right|}/{2}$ as a consequence of the half-infinite geometry. Using the definition \mbox{$V_{C}\left(\boldsymbol{q},z\right)=\intop d^{2}\boldsymbol{r}e^{-i\boldsymbol{q}\cdot\boldsymbol{r}}V_{C}\left(\boldsymbol{r},z\right)=\frac{2\pi C}{q}e^{-q\left|z\right|}$} and integrating over $\boldsymbol{R}$ and $\boldsymbol{r}$, we arrive at
\begin{eqnarray}
	H_C=\frac{1}{2\mathcal{S}^{3}}\sum_{\boldsymbol{q}}\sum_{\boldsymbol{k},\boldsymbol{k'}} \intop dZ\intop dz\psi_{\boldsymbol{k}+\boldsymbol{q}}^{\dagger}\left(Z+\frac{z}{2}\right)\psi_{\boldsymbol{k}'-\boldsymbol{q}}^{\dagger}\left(Z-\frac{z}{2}\right)V_{C}\left(\boldsymbol{q},z\right)\psi_{\boldsymbol{k}'}\left(Z-\frac{z}{2}\right)\psi_{\boldsymbol{k}}\left(Z+\frac{z}{2}\right)
\end{eqnarray} 
We derive now the effective interaction between Fermi arcs. To this end, we factorize the Fermi arc operator by isolating the $z$-dependent part of the wavefunction, see \eqref{eq:psitilde}. Performing the integrations in $Z$ and $z$, we obtain
\begin{equation}
	H_{C}=\intop\frac{d\boldsymbol{k}}{\left(2\pi\right)^{2}}\intop\frac{d\boldsymbol{k'}}{\left(2\pi\right)^{2}}\intop\frac{d\boldsymbol{q}}{\left(2\pi\right)^{2}}\mathcal{F}_{C}\left(\boldsymbol{k},\boldsymbol{k'},\boldsymbol{q}\right)c_{\boldsymbol{k}+\boldsymbol{q}}^{\dagger}c_{\boldsymbol{k}'-\boldsymbol{q}}^{\dagger}c_{\boldsymbol{k}'}c_{\boldsymbol{k}}
\end{equation}
with 
\begin{equation}\label{eq:CoulombFF-1}
	\mathcal{F}_{C}\left(\boldsymbol{k},\boldsymbol{k'},\boldsymbol{q}\right)=\frac{\sqrt{\kappa_{\boldsymbol{k}+\boldsymbol{q}}\kappa_{\boldsymbol{k}}\kappa_{\boldsymbol{k}'-\boldsymbol{q}}\kappa_{\boldsymbol{k}'}}}{\kappa_{\boldsymbol{k}+\boldsymbol{q}}+\kappa_{\boldsymbol{k}}+\kappa_{\boldsymbol{k}'-\boldsymbol{q}}+\kappa_{\boldsymbol{k}'}}\frac{8\pi C}{q}\left(\frac{1}{\kappa_{\boldsymbol{k}+\boldsymbol{q}}+\kappa_{\boldsymbol{k}}+q}+\frac{1}{\kappa_{\boldsymbol{k}'-\boldsymbol{q}}+\kappa_{\boldsymbol{k}'}+q}\right)\;.
\end{equation}
When specialized to the Cooper channel $\boldsymbol{k'}=-\boldsymbol{k}$ and to the pair operators $B_{\boldsymbol{k}}=\frac{1}{\sqrt{\mathcal{S}}}c_{-\boldsymbol{k}}c_{\boldsymbol{k}}$, we obtain (6) in the main text, with form factor
\begin{equation}\label{eq:FCkred}
	F_z\left(\boldsymbol{k},\boldsymbol{q}\right)=\frac{2\kappa_{\boldsymbol{k}+\boldsymbol{q}}\kappa_{\boldsymbol{k}}}{\left(\kappa_{\boldsymbol{k}+\boldsymbol{q}}+\kappa_{\boldsymbol{k}}\right)\left(\kappa_{\boldsymbol{k}+\boldsymbol{q}}+\kappa_{\boldsymbol{k}}+q\right)}
\end{equation}
This has the clear structure of the usual Coulomb potential modulated by an integral over the $z$ direction, encoding the wavevector-dependent penetration in the bulk. It is therefore an intrinsically three-dimensional object.

In the main text, we introduce a phenomenological screening wavevector $q_s$, see (6) in the main text. Its effect is illustrated in Fig. \ref{fig:FAopen}, depicting the full intra- and inter-arc interaction potentials. Decreasing the screening wavevector, the former turns repulsive on average, while the latter is only weakly affected. The Thomas-Fermi wavevector is a phenomenological parameter, in a way that it is convenient for use in the gap equations. In particular, eq. (6) in the main text should be understood as a static, long-wavelength and direction-averaged limit of a Coulomb potential in RPA 
\cite{Ghosh2020}.

\subsection{Electron-phonon interaction}\label{app:epinteractions}
Using the expansions
\begin{eqnarray}
	\boldsymbol{u}\left(\boldsymbol{r}\right)=\intop\frac{d\boldsymbol{q}}{\left(2\pi\right)^{2}}e^{i\boldsymbol{q}\cdot\boldsymbol{r}}\boldsymbol{u}\left(\boldsymbol{q},z\right) & \qquad & \psi\left(\boldsymbol{r}\right)=\intop\frac{d\boldsymbol{k}}{\left(2\pi\right)^{2}}e^{i\boldsymbol{k}\cdot\boldsymbol{r}}\psi_{\boldsymbol{k}}\left(z\right)\,,
\end{eqnarray}
the continuum limit of \eqref{eq:HepLattice} is obtained as 
\begin{eqnarray}
	H_{\text{ep}} & = & -U_{\text{ep}}\intop\frac{d\boldsymbol{k}}{\left(2\pi\right)^{2}}\intop\frac{d\boldsymbol{q}}{\left(2\pi\right)^{2}}\intop_{-\infty}^{0}dz\psi_{\boldsymbol{k}+\boldsymbol{q}}^{\dagger}\left\{ iq_{x}u_{x}\left(\boldsymbol{q},z\right)\sigma^{x}+iq_{y}u_{y}\left(\boldsymbol{q},z\right)\sigma^{y}+\left[\partial_{z}u_{z}\left(\boldsymbol{q},z\right)\right]\left(\sigma^{x}+\sigma^{y}\right)\right\} \psi_{\boldsymbol{k}}
\end{eqnarray}
We then substitute the mode expansion (\ref{eq:ufield-2-intdk})
in compact form, with the understanding that the mode index $\lambda$
is restricted by the frequency. Collecting the various terms
\begin{eqnarray}
	iq_{j}u_{j}\left(\boldsymbol{q}\right)+\partial_{z}u_{z}\left(\boldsymbol{q}\right) & = & \frac{1}{2}\sqrt{\frac{\hbar}{2\omega\rho}}\sum_{\lambda,\omega}\Bigg\{ a_{\omega,\boldsymbol{q}}^{(\lambda)}\left(t\right)\left[iq_{j}w_{\omega,\boldsymbol{q}}^{(\lambda,j)}\left(z\right)+\partial_{z}w_{\omega,\boldsymbol{q}}^{(\lambda,z)}\left(z\right)\right]\nonumber \\
	&  & \qquad\qquad+a_{\omega,-\boldsymbol{q}}^{(\lambda)\dagger}\left(t\right)\left[iq_{j}w_{\omega,-\boldsymbol{q}}^{(\lambda,j)*}\left(z\right)+\partial_{z}w_{\omega,-\boldsymbol{q}}^{(\lambda,z)*}\left(z\right)\right]\Bigg\}
\end{eqnarray}
Factor now the $z$ dependence of the fermionic fields as
\begin{equation}
	\psi_{\boldsymbol{k}}\left(z\right)=\tilde{\psi}_{\boldsymbol{k}}\sqrt{2\kappa_{\boldsymbol{k}}}e^{\kappa_{\boldsymbol{k}}z}\qquad\qquad\tilde{\psi}_{\boldsymbol{k}}=\chi_{-}c_{\boldsymbol{k}}\label{eq:psitilde}
\end{equation}
i.e., in $\tilde{\psi}_{\boldsymbol{k}}$ there remain the ladder
operators $c_{\boldsymbol{k}}$ creating a spinor with fixed orientation,
as well as the spinorial structure in Eq. \eqref{eq:xipm}.
Conversely, the exponential part of the wavefunction and its normalization
are explicitly written. Then we can cast the electron-phonon interaction
in the form
\begin{eqnarray}
	H_{\text{ep}} & = & \sum_{j=x,y}\sum_{\lambda}\intop\frac{d\boldsymbol{k}}{\left(2\pi\right)^{2}}\intop\frac{d\boldsymbol{q}}{\left(2\pi\right)^{2}}\sum_{\omega}\left[g_{\lambda,\boldsymbol{k},\boldsymbol{q},\omega}^{(j)}a_{\omega,\boldsymbol{q}}^{(\lambda)}\left(t\right)+g_{\lambda,\boldsymbol{k},\boldsymbol{q},\omega}^{(j)*}a_{\omega,-\boldsymbol{q}}^{(\lambda)\dagger}\left(t\right)\right]\tilde{\psi}_{\boldsymbol{k}+\boldsymbol{q}}^{\dagger}\sigma^{j}\tilde{\psi}_{\boldsymbol{k}}+\text{H.c.}\label{eq:HepC}
\end{eqnarray}
with a sum over the phonon mode index $\lambda$ and the corresponding
directional coupling
\begin{eqnarray}
	g_{\lambda,\boldsymbol{k},\boldsymbol{q},\omega}^{(j)} & = & U_{\text{ep}}\sqrt{\frac{\hbar}{2\omega\rho}}\sqrt{\kappa_{\boldsymbol{k}}\kappa_{\boldsymbol{k}+\boldsymbol{q}}}\intop_{-\infty}^{0}dz\left[iq_{j}w_{\omega,\boldsymbol{q}}^{(\lambda,j)}\left(z\right)+\partial_{z}w_{\omega,\boldsymbol{q}}^{(\lambda,z)}\left(z\right)\right]e^{\left[\kappa_{}\left(\boldsymbol{k}\right)+\kappa_{}\left(\boldsymbol{k}+\boldsymbol{q}\right)\right]z}\label{eq:gdef}
\end{eqnarray}
The first term in (\ref{eq:HepC}) models the transition from a state
with momentum $\boldsymbol{k}$ to one at momentum $\boldsymbol{k}+\boldsymbol{q}$
via absorption of a phonon with momentum $\boldsymbol{q}$, the second
via emission of a phonon with momentum $-\boldsymbol{q}$. Other types
of coupling are in principle allowed by symmetry \cite{Pereira2019},
but involve higher derivatives of the phonon modes or of the electronic
modes. These will not be included in the model, as they would produce
subleading corrections to the temperature dependence in the low-temperature
regime \cite{Buccheri2022}.
Using the contraction
\begin{equation}\label{eq:contraction}
	\mbox{$\tilde{\psi}_{\boldsymbol{k'}}^{\dagger}\sigma^{x}\tilde{\psi}_{\boldsymbol{k}} = - \cos\alpha c_{\boldsymbol{k'}}^{\dagger}c_{\boldsymbol{k}}$}\;, \qquad \mbox{$
		\tilde{\psi}_{\boldsymbol{k'}}^{\dagger}\sigma^{y}\tilde{\psi}_{\boldsymbol{k}}=-\sin\alpha\,c_{\boldsymbol{k'}}^{\dagger}c_{\boldsymbol{k}}$} \;, \qquad \mbox{$\tilde{\psi}_{\boldsymbol{k'}}^{\dagger}\sigma^{z}\tilde{\psi}_{\boldsymbol{k}}=0$}
\end{equation}
we arrive at the interaction (4) in the main text, with
$ g_{\lambda,\boldsymbol{k},\boldsymbol{q},\omega}=g_{\lambda,\boldsymbol{k},\boldsymbol{q},\omega}^{(x)}\cos\alpha+g_{\lambda,\boldsymbol{k},\boldsymbol{q},\omega}^{(y)}\sin\alpha$.

\subsection{Phonon-mediated electron-electron interaction}\label{sec:Phonon-mediated-ee}

We now derive the effective electron-electron interaction Hamiltonian
by second-order perturbation theory.
We retain only surface states as initial and final states, so the energy
will not appear as an independent quantum number and the initial state will be labeled by the pair of momenta $\left|\boldsymbol{k};\boldsymbol{k'}\right\rangle$.
One phonon is then exchanged between the electrons and the final states
are written as $\left|\boldsymbol{k}+\boldsymbol{q};\boldsymbol{k'}-\boldsymbol{q}\right\rangle $.
We write the matrix elements of $H^{(2)}$ between these states
using a Schrieffer-Wolff transformation. We look for a similarity transformation in the form
\begin{eqnarray}
	H' & = & e^{-\eta S}He^{\eta S}\approx H+\eta\left[H,S\right]+\frac{\eta^{2}}{2}\left[\left[H,S\right],S\right]+\mathcal{O}\left(\eta^{3}\right)\label{eq:HprimeSW}
\end{eqnarray}
where we assume that $\eta\sim V_{ep}/\hbar \omega_{q}\ll1$. For acoustic branches this is not a uniformly controlled approximation and may break down for $q\to0$. Nevertheless, the small-momentum scattering is dominated by the Coulomb repulsion, which effectively introduces a cutoff. 
In order to eliminate $H_{ep}$ we require that the generator $S$
satisfies $H_{ep}=-\left[H_{0},S\right]$ and $S=-S^{\dagger}$. We
employ the Ansatz
\begin{eqnarray}
	S & = & \sum_{j=x,y}\sum_{\lambda}\intop\frac{d\boldsymbol{k}}{\left(2\pi\right)^{2}}\intop\frac{d\boldsymbol{q}}{\left(2\pi\right)^{2}}\sum_{\omega}\left[f_{\lambda,\boldsymbol{k},\boldsymbol{q},\omega}^{(j)}a_{\omega,\boldsymbol{q}}^{(\lambda)}+h_{\lambda,\boldsymbol{k},\boldsymbol{q},\omega}^{(j)*}a_{\omega,-\boldsymbol{q}}^{(\lambda)\dagger}\right]\tilde{\psi}_{\boldsymbol{k}+\boldsymbol{q}}^{\dagger}\sigma^{j}\tilde{\psi}_{\boldsymbol{k}}-\mbox{H.c.}\label{eq:SWgenAnsatz}
\end{eqnarray}
and we determine the still unknown scalar functions $f_{\lambda,\boldsymbol{k},\boldsymbol{q},\omega}^{(j)}$
and $h_{\lambda,\boldsymbol{k},\boldsymbol{q},\omega}^{(j)}$ by explicitly
computing the commutators with $H_{0}=H_{0,e}+H_{0,p}$. We write the SW generator in
the form
\begin{eqnarray}
	S & = & \sum_{j=x,y}\sum_{\lambda}\intop\frac{d\boldsymbol{k}}{\left(2\pi\right)^{2}}\intop\frac{d\boldsymbol{q}}{\left(2\pi\right)^{2}}\sum_{\omega}\left[\frac{g_{\lambda,\boldsymbol{k},\boldsymbol{q},\omega}^{(j)}a_{\omega,\boldsymbol{q}}^{(\lambda)}}{\zeta\left(\boldsymbol{k},\boldsymbol{q}\right)+\hbar\omega}+\frac{g_{\lambda,\boldsymbol{k},\boldsymbol{q},\omega}^{(j)*}a_{\omega,-\boldsymbol{q}}^{(\lambda)\dagger}}{\zeta\left(\boldsymbol{k},\boldsymbol{q}\right)-\hbar\omega}\right]\tilde{\psi}_{\boldsymbol{k}+\boldsymbol{q}}^{\dagger}\sigma^{j}\tilde{\psi}_{\boldsymbol{k}}-\mbox{H.c.}\label{eq:SWgenerator}
\end{eqnarray}
where $\zeta(\boldsymbol{k},\boldsymbol{q})=\varepsilon_{\boldsymbol{k}+\boldsymbol{q}}-\varepsilon_{\boldsymbol{k}}$. 
From this expression, we compute the effective
phonon-mediated electron-electron interaction as the first non-vanishing
contribution in the expansion (\ref{eq:HprimeSW}). This is at $\mathcal{O}\left(\eta^{2}\right)$
and is the sum of the remaining term in $-\eta\left[S,H_{ep}\right]$
and the following one in (\ref{eq:HprimeSW}). In short, we need the
commutator
\begin{eqnarray}
	H^{(2)} & = & -\frac{1}{2}\left[S,H_{ep}\right]\,=\,-\frac{1}{2}\sum_{j,l=x,y}\sum_{\lambda,\mu}\intop\frac{d\boldsymbol{k}}{\left(2\pi\right)^{2}}\intop\frac{d\boldsymbol{q}}{\left(2\pi\right)^{2}}\sum_{\omega}\intop\frac{d\boldsymbol{k}_{1}}{\left(2\pi\right)^{2}}\intop\frac{d\boldsymbol{q}_{1}}{\left(2\pi\right)^{2}}\sum_{\omega_{1}}\\
	&  & \Bigg[\left(\frac{g_{\lambda,\boldsymbol{k},\boldsymbol{q},\omega}^{(j)}a_{\omega,\boldsymbol{q}}^{(\lambda)}}{\zeta\left(\boldsymbol{k},\boldsymbol{q}\right)+\hbar\omega}+\frac{g_{\lambda,\boldsymbol{k},\boldsymbol{q},\omega}^{(j)*}a_{\omega,-\boldsymbol{q}}^{(\lambda)\dagger}}{\zeta\left(\boldsymbol{k},\boldsymbol{q}\right)-\hbar\omega}\right)\tilde{\psi}_{\boldsymbol{k}+\boldsymbol{q}}^{\dagger}\sigma^{j}\tilde{\psi}_{\boldsymbol{k}}-\left(\frac{g_{\lambda,\boldsymbol{k},\boldsymbol{q},\omega}^{(j)*}a_{\omega,\boldsymbol{q}}^{(\lambda)\dagger}}{\zeta\left(\boldsymbol{k},\boldsymbol{q}\right)+\hbar\omega}+\frac{g_{\lambda,\boldsymbol{k},\boldsymbol{q},\omega}^{(j)}a_{\omega,-\boldsymbol{q}}^{(\lambda)}}{\zeta\left(\boldsymbol{k},\boldsymbol{q}\right)-\hbar\omega}\right)\tilde{\psi}_{\boldsymbol{k}}^{\dagger}\sigma^{j}\tilde{\psi}_{\boldsymbol{k}+\boldsymbol{q}},\nonumber \\
	&  & \left(g_{\mu,\boldsymbol{k}_{1},\boldsymbol{q}_{1},\omega_{1}}^{(l)}a_{\omega_{1},\boldsymbol{q}_{1}}^{(\mu)}+g_{\mu,\boldsymbol{k}_{1},\boldsymbol{q}_{1},\omega_{1}}^{(l)*}a_{\omega_{1},-\boldsymbol{q}_{1}}^{(\mu)\dagger}\right)\tilde{\psi}_{\boldsymbol{k}_{1}+\boldsymbol{q}_{1}}^{\dagger}\sigma^{l}\tilde{\psi}_{\boldsymbol{k}_{1}}+\left(g_{\mu,\boldsymbol{k}_{1},\boldsymbol{q}_{1},\omega_{1}}^{(l)*}a_{\omega_{1},\boldsymbol{q}_{1}}^{(\mu)\dagger}+g_{\mu,\boldsymbol{k}_{1},\boldsymbol{q}_{1},\omega_{1}}^{(l)}a_{\omega_{1},-\boldsymbol{q}_{1}}^{(\mu)}\right)\tilde{\psi}_{\boldsymbol{k}_{1}}^{\dagger}\sigma^{l}\tilde{\psi}_{\boldsymbol{k}_{1}+\boldsymbol{q}_{1}}\Bigg]\nonumber 
\end{eqnarray}
We eventually arrive at a phonon-mediated electron-electron interaction in the form
\begin{equation}
	H_{ee}  =  -\frac{1}{2}\sum_{j,l=x,y}\sum_{\lambda}\intop\frac{d\boldsymbol{k}d\boldsymbol{k}_{1}d\boldsymbol{q}}{\left(2\pi\right)^{6}}\sum_{\omega}\left[\frac{G_{\lambda,\boldsymbol{k},\boldsymbol{q},\omega}^{(j)}G_{\lambda,\boldsymbol{k}_{1},-\boldsymbol{q},\omega}^{(l)*}}{\zeta\left(\boldsymbol{k},\boldsymbol{q}\right)+\hbar\omega}-\frac{G_{\lambda,\boldsymbol{k},\boldsymbol{q},\omega}^{(j)*}G_{\lambda,\boldsymbol{k}_{1},-\boldsymbol{q},\omega}^{(l)}}{\zeta\left(\boldsymbol{k},\boldsymbol{q}\right)-\hbar\omega}\right]\tilde{\psi}_{\boldsymbol{k}+\boldsymbol{q}}^{\dagger}\sigma^{j}\tilde{\psi}_{\boldsymbol{k}}\tilde{\psi}_{\boldsymbol{k}_{1}-\boldsymbol{q}}^{\dagger}\sigma^{l}\tilde{\psi}_{\boldsymbol{k}_{1}}
	\label{eq:Heep}
\end{equation}
where
\begin{equation}
	G_{\lambda,\boldsymbol{k},\boldsymbol{q},\omega}^{(j)} =  g_{\lambda,\boldsymbol{k},\boldsymbol{q},\omega}^{(j)}+g_{\lambda,\boldsymbol{k}+\boldsymbol{q},-\boldsymbol{q},\omega}^{(j)}\,=\,G_{\lambda,\boldsymbol{k}+\boldsymbol{q},-\boldsymbol{q},\omega}^{(j)}\;.\label{eq:G=00003Dg+g}
\end{equation}
Finally, we note that the remaining terms in the
commutator $\left[S,H_{ep}\right]$ can be gathered in a new electron-phonon
interaction of the form
\begin{equation}\label{eq:Heppe1}
	H_{epp} =  \sum_{j=x,y}\sum_{\lambda}\intop\frac{d\boldsymbol{k}}{\left(2\pi\right)^{2}}\intop\frac{d\boldsymbol{q}}{\left(2\pi\right)^{2}}\intop\frac{d\boldsymbol{q}_{0}}{\left(2\pi\right)^{2}}\sum_{\omega}\,f_{\lambda,\boldsymbol{k},\boldsymbol{q}_{0},\boldsymbol{q},\omega}^{(j)}a_{\omega,\boldsymbol{q}_{0}-\boldsymbol{q}}^{(\lambda)\dagger}a_{\omega,\boldsymbol{q}_{0}+\boldsymbol{q}}^{(\lambda)}\tilde{\psi}_{\boldsymbol{k}+\boldsymbol{q}}^{\dagger}\sigma^{j}\tilde{\psi}_{\boldsymbol{k}}\;,
\end{equation}
for some functions $f_{\lambda,\boldsymbol{k},\boldsymbol{q}_{0},\boldsymbol{q},\omega}^{(j)}$. At second order, this also generates a quartic electron-electron interaction
of the form \eqref{eq:Heep}. As such terms simply renormalize the interaction
found before, they will not be explicitly included in what follows.
The fermions have fixed orientation on the surface, see \eqref{eq:xipm}. Using \eqref{eq:contraction}, we can perform the sum over $j$, $l$ in \eqref{eq:Heep} and arrive to
\begin{equation}
	H_{ee}  = - \frac{1}{2}\sum_{\lambda}\intop\frac{d\boldsymbol{k}d\boldsymbol{k}_{1}d\boldsymbol{q}}{\left(2\pi\right)^{6}}\sum_{\omega}\left[\frac{G_{\lambda,\boldsymbol{k},\boldsymbol{q},\omega}G_{\lambda,\boldsymbol{k}_{1},-\boldsymbol{q},\omega}^{*}}{\zeta\left(\boldsymbol{k},\boldsymbol{q}\right)+\hbar\omega}-\frac{G_{\lambda,\boldsymbol{k},\boldsymbol{q},\omega}^{*}G_{\lambda,\boldsymbol{k}_{1},-\boldsymbol{q},\omega}}{\zeta\left(\boldsymbol{k},\boldsymbol{q}\right)-\hbar\omega}\right]c_{\boldsymbol{k}+\boldsymbol{q}}^{\dagger}c_{\boldsymbol{k}_{1}-\boldsymbol{q}}^{\dagger}c_{\boldsymbol{k}_{1}}c_{\boldsymbol{k}}
\end{equation}
In our formalism  $c_{\boldsymbol{k}}^{\dagger}$ creates an electron in a Fermi arc state, with 
a specific orientation of the internal degree of freedom. Phonons do not couple to such degree of freedom, which is why the interaction \eqref{eq:Heep} can be cast into the simpler form (5) in the main text.

\section{Computation of the phonon potentials}\label{app:potentials}


In this Appendix, we compute the matrix elements involving the various families of phonon modes, exchanged between a pair of electrons on the surface. 
At fixed momentum, the largest matrix elements come from the R phonons. These vibrations have a component along the wavevector $\boldsymbol{q}$ and are exponentially damped along $z$, see \eqref{eq:Rayleighx} and \eqref{eq:NRx}. In particular
\begin{equation}
	iq_{x/y}w_{\boldsymbol{q}}^{(R,x/y)}\left(z\right)+\partial_{z}w_{\boldsymbol{q}}^{(R,z)}\left(z\right) =  \frac{i}{q}\mathcal{N}_{\boldsymbol{q}}^{(R)}\left[\frac{\left(\xi_{R}^{2}\nu^{2}q^{2}-q_{y/x}^{2}\right)\left(2-\xi_{R}^{2}\right)}{2\sqrt{1-\xi_{R}^{2}\nu^{2}}}e^{\sqrt{1-\xi_{R}^{2}\nu^{2}}qz}+q_{x}^{2}\sqrt{1-\xi_{R}^{2}}e^{\sqrt{1-\xi_{R}^{2}}qz}\right] 
\end{equation}
Computing the integral over $z$
\begin{eqnarray}
	\intop_{-\infty}^{0}dze^{\left[\kappa_{l,t}+\kappa_{}\left(\boldsymbol{k}\right)+\kappa_{}\left(\boldsymbol{k}+\boldsymbol{q}\right)\right]z} & = & \frac{1}{\kappa_{l,t}+\kappa_{\boldsymbol{k}}+\kappa_{\boldsymbol{k}+\boldsymbol{q}}}
\end{eqnarray}
and using \eqref{eq:NRx}, together with the properties $g_{R,\boldsymbol{k},\boldsymbol{q}}^{(x/y)}=g_{R,\boldsymbol{k}+\boldsymbol{q},-\boldsymbol{q}}^{(x/y)}$ and  $G_{R,\boldsymbol{k},\boldsymbol{q}}^{(x/y)}=2g_{R,\boldsymbol{k},\boldsymbol{q}}^{(x/y)}$, we arrive at the vertex
\begin{equation}
	G_{R,\boldsymbol{k},\boldsymbol{q}}=iU_{\text{ep}}\sqrt{\frac{1}{\rho}}\frac{\mathcal{N}_{0}^{(R)}}{\sqrt{\xi_{R}c_{t}}q}\left\{ \frac{\left(2-\xi_{R}^{2}\right)\left[\left(\xi_{R}^{2}\nu^{2}q^{2}-q_{y}^{2}\right)\cos\alpha+\left(\xi_{R}^{2}\nu^{2}q^{2}-q_{x}^{2}\right)\sin\alpha\right]}{\sqrt{1-\xi_{R}^{2}\nu^{2}}\left[\sqrt{1-\xi_{R}^{2}\nu^{2}}q+\kappa_{\boldsymbol{k}}+\kappa_{\boldsymbol{k}+\boldsymbol{q}}\right]}+\frac{2\left(q_{y}^{2}\cos\alpha+q_{x}^{2}\sin\alpha\right)\sqrt{1-\xi_{R}^{2}}}{\sqrt{1-\xi_{R}^{2}}q+\kappa_{\boldsymbol{k}}+\kappa_{\boldsymbol{k}+\boldsymbol{q}}}\right\} \label{eq:GsqR}
\end{equation}

One can proceed similarly for $t$ modes (\ref{eq:transverse}), obtaining
\begin{equation}
	\left|G_{t,\boldsymbol{k},\boldsymbol{q},\omega}^{(j)}\right|^{2}  =  \frac{\hbar U_{\text{ep}}^{2}}{\rho L}\frac{1}{\omega}\frac{16q_{l}^{2}}{\omega_{t}^{4}\omega_{l}^{2}-4q^{4}\left(\omega_{t}^{2}-\omega_{l}^{2}\right)}\left\{ \frac{\left(q^{2}-q_{t}^{2}\right)\left[\left(q_{x}^{2}+q_{l}^{2}\right)\cos\alpha+\left(q_{y}^{2}+q_{l}^{2}\right)\sin\alpha\right]}{2\left[\left(\kappa_{\boldsymbol{k}}+\kappa_{\boldsymbol{k}+\boldsymbol{q}}\right)^{2}+q_{l}^{2}\right]}-\frac{\left(q_{y}^{2}\cos\alpha+q_{x}^{2}\sin\alpha\right)q_{t}^{2}}{\left[\left(\kappa_{\boldsymbol{k}}+\kappa_{\boldsymbol{k}+\boldsymbol{q}}\right)^{2}+q_{t}^{2}\right]}\right\} ^{2}\label{eq:Gsqt}
\end{equation}
The corresponding contribution to the potential 
\begin{eqnarray}\label{eq:GtoV}
	V_{t}\left(\boldsymbol{k},\boldsymbol{q}\right) =2\kappa_{}\left(\boldsymbol{k}+\boldsymbol{q}\right)\kappa_{}\left(\boldsymbol{k}\right)\intop d\omega\left|G_{t,\boldsymbol{k},\boldsymbol{q},\omega}^{(j)}\right|^{2}\left(\frac{1}{\zeta\left(\boldsymbol{k},\boldsymbol{q}\right)-\omega}-\frac{1}{\zeta\left(\boldsymbol{k},\boldsymbol{q}\right)+\omega}\right) \;,
\end{eqnarray}
is computed numerically.

For l modes (\ref{eq:longitudinal}), we obtain in the same way
\begin{equation}
	\left|G_{l,\boldsymbol{k},\boldsymbol{q},\omega}\right|^{2}  = 
	\frac{\hbar}{\rho L}U_{\text{ep}}^{2}\frac{1}{\omega}\frac{16q^{2}q_{t}^{2}\left(\kappa_{\boldsymbol{k}}+\kappa_{\boldsymbol{k}+\boldsymbol{q}}\right)^{2}}{\omega_{t}^{6}-4q^{2}q_{t}^{2}\left(\omega_{t}^{2}-\omega_{l}^{2}\right)}\left\{ \frac{\left[\left(\omega_{l}^{2}-q_{y}^{2}\right)\cos\alpha+\left(\omega_{l}^{2}-q_{x}^{2}\right)\sin\alpha\right]}{\left(\kappa_{\boldsymbol{k}}+\kappa_{\boldsymbol{k}+\boldsymbol{q}}\right)^{2}+q_{l}^{2}}-\frac{\omega_{t}^{2}-2q^{2}}{2q^{2}}\frac{\left(q_{y}^{2}\cos\alpha+q_{x}^{2}\sin\alpha\right)}{\left(\kappa_{\boldsymbol{k}}+\kappa_{\boldsymbol{k}+\boldsymbol{q}}\right)^{2}+q_{t}^{2}}\right\} ^{2}\label{eq:Gsql}
\end{equation}

Shear modes \eqref{eq:shear} and $m$ modes From \eqref{eq:mixed}  carry a negligible weight compared to the other contribution. We write their matrix elements as
\begin{equation}\label{eq:Gsqs}
	G_{s,\boldsymbol{k},\boldsymbol{q},\omega}  =  -i\sqrt{\frac{\hbar}{\rho L\omega}}U_{\text{ep}}\frac{8q_{x}q_{y}}{q}\frac{\left(\kappa_{\boldsymbol{k}}+\kappa_{\boldsymbol{k}+\boldsymbol{q}}\right)\left(\cos\alpha-\sin\alpha\right)}{\left(\kappa_{\boldsymbol{k}}+\kappa_{\boldsymbol{k}+\boldsymbol{q}}\right)^{2}+q_{t}^{2}}
\end{equation}
and
\begin{eqnarray}\label{eq:Gsqm}
	\left|G_{m,\boldsymbol{k},\boldsymbol{q},\omega}\right|^{2} & = & \frac{\hbar}{L\rho}U_{\text{ep}}^{2}\frac{1}{\omega}\frac{64\kappa_{l}^{2}q_{t}^{2}q^{2}}{\omega_{t}^{2}\left[16\kappa_{l}^{2}q_{t}^{2}q^{4}+\left(q_{t}^{2}-q^{2}\right)^{4}\right]}\Bigg[\frac{\left(q_{t}^{2}-q^{2}\right)\left[\left(q_{y}^{2}-\omega_{l}^{2}\right)\cos\alpha+\left(q_{x}^{2}-\omega_{l}^{2}\right)\sin\alpha\right]}{2\kappa_{l}\left(\kappa_{\boldsymbol{k}}+\kappa_{\boldsymbol{k}+\boldsymbol{q}}+\kappa_{l}\right)}\nonumber \\
	&  & \qquad\qquad\qquad-\frac{q_{y}^{2}\cos\alpha+q_{x}^{2}\sin\alpha}{\left(\kappa_{\boldsymbol{k}}+\kappa_{\boldsymbol{k}+\boldsymbol{q}}\right)^{2}+q_{t}^{2}}\left(q_{t}^{2}-\frac{\left(q_{t}^{2}-q^{2}\right)^{2}\left(\kappa_{\boldsymbol{k}}+\kappa_{\boldsymbol{k}+\boldsymbol{q}}\right)}{4q^{2}\kappa_{l}}\right)\Bigg]^{2}.
\end{eqnarray}
We underline in all these expressions the factors $1/L$, scaling like the inverse of the transverse size.

\section{BdG formalism}\label{app:gap}
\subsection{Derivation of the gap equation}
We aim at the momentum-dependence of the mean-field gap function, satisfying (8) in the main text. The effective Hamiltonian on the arcs is written as
\begin{eqnarray}\label{eq:1dHamiltonian}
	H&=&\sum_{\tau}\intop d\varepsilon\intop\frac{dk_{x}}{2\pi}\Bigg\{\varepsilon n_{}\left(\varepsilon,k_{x}\right)c_{\tau,k_{x}}^{\dagger}c_{\tau,k_{x}}+\sum_{\tau',\tau_{1}}\intop d\varepsilon'\intop\frac{dk'_{x}}{2\pi}\intop d\varepsilon_{1}\intop\frac{dk_{1,x}}{2\pi}n_{}\left(\varepsilon,k_{x}\right)n_{}\left(\varepsilon',k'_{x}\right)n_{}\left(\varepsilon_{1},k_{1,x}\right)\\&&\qquad\qquad\qquad\qquad\qquad\times\mathcal{V}_{\tau',\tau,\tau_{1}}\left(\varepsilon',k'_{x};\varepsilon,k_{x};\varepsilon,k_{1,x}\right)c_{\tau_{1}-\tau'+\tau,\varepsilon+\varepsilon_{1}-\varepsilon',k_{1,x}-k'_{x}+k_{x}}^{\dagger}c_{\tau',\varepsilon',k'_{x}}^{\dagger}c_{\tau,\varepsilon,k_{x}}c_{\tau_{1},\varepsilon_{1},k_{1,x}}\Bigg\}\nonumber,
\end{eqnarray}
in which a generic two-body interaction potential 
$V(\boldsymbol{k},\boldsymbol{k}',\boldsymbol{k}_1)$ has been rewritten using the variables $(\varepsilon,\tau,k_x)$ for each of the arguments. When not ambiguous, we will sometimes omit the argument $\varepsilon$ for lightness of notation. We proceed with the mean-field decomposition: define the dimensionless anomalous expectation value $F\left(\boldsymbol{k}\right)$ via the relation $\left\langle c_{\boldsymbol{k}}c_{\boldsymbol{k}_{1}}\right\rangle \equiv\left(2\pi\right)^{2}\delta^{(2)}\left(\boldsymbol{k}+\boldsymbol{k}_{1}\right)F\left(\boldsymbol{k}\right)$. The change of variables $\left(k_{x},k_{y}\right)\to\left(\varepsilon,\tau,k_{x}\right)$ is performed via \eqref{eq:ky} and implies that $\delta\left(k_{y}+k_{1,y}\right)=\left[2\pi n_{}\left(\varepsilon,k_{x}\right)\right]^{-1}\delta\left(\varepsilon_{1}-\varepsilon\right)$. This relation is employed in writing the anomalous expectation value as
\begin{equation}
	\left\langle c_{\tau,\varepsilon,k_{x}}c_{\tau_{1},\varepsilon_{1},k_{1,x}}\right\rangle =2\pi\delta_{\tau_{1},-\tau}\delta\left(k_{x}+k_{1,x}\right)\left[n_{}\left(\varepsilon,k_{x}\right)\right]^{-1}\delta\left(\varepsilon_{1}-\varepsilon\right)F_{\tau}\left(\varepsilon,k_{x}\right)
\end{equation}
and in integrals of the form
\begin{equation}
	\intop\frac{d^{2}\boldsymbol{k}}{\left(2\pi\right)^{2}}\intop\frac{d^{2}\boldsymbol{k}'}{\left(2\pi\right)^{2}}\left(2\pi\right)^{2}\delta^{(2)}\left(\boldsymbol{k}+\boldsymbol{k'}\right)\ldots
	= \sum_{\tau}\intop d\varepsilon\intop\frac{dk_{x}}{2\pi}n_{}\left(\varepsilon,k_{x}\right)\ldots
\end{equation}
Note that 
\begin{equation}\label{eq:antisymmF}
	F_{-}\left(\varepsilon,-k_{x}\right)=-F_{+}\left(\varepsilon,k_{x}\right)
\end{equation}
because of the fermionic statistics. Then we approximate the interaction terms in the Cooper channel as
\begin{align}
	&c_{\tau_{1}-\tau'+\tau,k_{1,x}-q_{x}}^{\dagger}c_{\tau',k_{x}+q_{x}}^{\dagger}c_{\tau,k_{x}}c_{\tau_{1},k_{1,x}} \\
	&\qquad= 2\pi\delta_{\tau_{1},-\tau}\delta\left(k_{x}+k_{1,x}\right)\left[n_{}\left(\varepsilon,k_{x}\right)\right]^{-1}\delta\left(\varepsilon_{1}-\varepsilon\right)\left[c_{-\tau',-k'_{x}}^{\dagger}c_{\tau',k'_{x}}^{\dagger}F_{\tau}\left(\varepsilon,k_{x}\right)+F_{\tau}^{*}\left(\varepsilon,k'_{x}\right)c_{\tau,k_{x}}c_{-\tau,-k_{x}}\right] \,,
	\nonumber
\end{align}
which brings the Hamiltonian \eqref{eq:1dHamiltonian} in the form
\begin{equation}\label{eq:Hintek}
	H=\sum_{\tau,\tau'}\intop d\varepsilon\intop\frac{dk_{x}}{2\pi}\intop d\varepsilon'\intop\frac{dk'_{x}}{2\pi}n_{}\left(\varepsilon,k_{x}\right)n_{}\left(\varepsilon',k'_{x}\right)\mathcal{V}_{\tau',\tau}\left(\varepsilon',k'_{x};\varepsilon,k_{x}\right)\left[c_{-\tau',-k'_{x}}^{\dagger}c_{\tau',k'_{x}}^{\dagger}F_{\tau}\left(\varepsilon,k_{x}\right)+F_{\tau'}^{*}\left(\varepsilon,k'_{x}\right)c_{\tau,k_{x}}c_{-\tau,-k_{x}}\right]
\end{equation}
where $\mathcal{V}_{\tau',\tau}\left(\varepsilon',k'_{x};\varepsilon,k_{x}\right)=\mathcal{V}_{\tau',\tau}\left(\varepsilon',k'_{x};\varepsilon,k_{x};\varepsilon,-k_{x}\right)$. 
In the new variables, the full Hamiltonian becomes an integral over energy shells $H=\intop d\varepsilon H_{\varepsilon}=\sum_{\tau}\intop d\varepsilon\intop\frac{dk_{x}}{2\pi}H_{\tau,\varepsilon,k_{x}}$. Let us now write a gap equation, starting from \eqref{eq:Hintek}. Define
\begin{equation}\label{eq:Delta(e,kx)}
	\frac{1}{2}\Delta_{\tau}\left(\varepsilon;k_{x}\right)=\sum_{\tau'}\intop d\varepsilon'\intop\frac{dk'_{x}}{2\pi}n_{}\left(\varepsilon',k'_{x}\right)\mathcal{V}_{\tau,\tau'}\left(\varepsilon,k_{x};\varepsilon',k'_{x}\right)F_{\tau'}\left(\varepsilon',k'_{x}\right)
\end{equation}
so that the shell Hamiltonian is written as
\begin{equation}\label{eq:Htekxodd}
	H_{\tau,\varepsilon,k_{x}}=\frac{1}{2}\varepsilon n_{}\left(\varepsilon,k_{x}\right)\left(c_{\tau,k_{x}}^{\dagger}c_{\tau,k_{x}}-c_{\tau,k_{x}}c_{\tau,k_{x}}^{\dagger}\right)-\frac{1}{2}n_{}\left(\varepsilon,k_{x}\right)\Delta_{\tau}\left(\varepsilon;k_{x}\right)c_{\tau,k_{x}}^{\dagger}c_{-\tau,-k_{x}}^{\dagger}-\frac{1}{2}n_{}\left(\varepsilon,k_{x}\right)\Delta_{\tau}^{*}\left(\varepsilon;k_{x}\right)c_{-\tau,-k_{x}}c_{\tau,k_{x}}
\end{equation}
The definition \eqref{eq:Delta(e,kx)} already contains an hypothesis on the sign structure of the gap in opposite quadrants of the BZ. To see this, let us define symmetric and antisymmetric combinations
\begin{equation}\label{eq:Deltaeo}
	\Delta_{e/o}\left(\varepsilon;k_{x}\right)\equiv\frac{\Delta_{+}\left(\varepsilon;k_{x}\right)\pm\Delta_{-}\left(\varepsilon;-k_{x}\right)}{2}
\end{equation}
Using \eqref{eq:Delta(e,kx)} and \eqref{eq:Deltaeo} and the fermionic statistics encoded in \eqref{eq:antisymmF}, we see that $	\Delta_{e}$ identically vanishes, which implies the sign structure
\begin{equation}\label{eq:Delta+(kx)=-Delta-(-kx)}
	\Delta_{+}\left(\varepsilon;k_{x}\right)=-\Delta_{-}\left(\varepsilon;-k_{x}\right)
\end{equation}
The antisymmetrized counterpart satisfies the equation
\begin{equation}\label{eq:Deltaodd}
	\Delta_{o}\left(\varepsilon;k_{x}\right)=2\intop d\varepsilon'\intop\frac{dk'_{x}}{2\pi}n_{}\left(\varepsilon',k'_{x}\right)\left\{ \mathcal{V}_{+,+}\left(\varepsilon,k_{x};\varepsilon',k'_{x}\right)F_{+}\left(\varepsilon',k'_{x}\right)+\mathcal{V}_{+,-}\left(\varepsilon,k_{x};\varepsilon',k'_{x}\right)F_{-}\left(\varepsilon',k'_{x}\right)\right\} 
\end{equation}
Writing the sum over the pairing terms in \eqref{eq:Htekxodd}, one obtains the pairing Hamiltonian 
\begin{equation}
	H_{\Delta}=-\intop d\varepsilon\intop\frac{dk_{x}}{2\pi}n_{}\left(\varepsilon,k_{x}\right)\left[\Delta_{o}\left(\varepsilon;k_{x}\right)c_{+,k_{x}}^{\dagger}c_{-,-k_{x}}^{\dagger}+\Delta_{o}^{*}\left(\varepsilon;k_{x}\right)c_{-,-k_{x}}c_{+,k_{x}}\right]
\end{equation}
which only contains the antisymmetric part. In alternative to \eqref{eq:Delta(e,kx)}, one can also define
\begin{equation}\label{eq:Delta(e,kx)other}
	\frac{1}{2}\Delta_{\tau}\left(\varepsilon;k_{x}\right)=\tau\sum_{\tau'}\intop d\varepsilon'\intop\frac{dk'_{x}}{2\pi}n_{}\left(\varepsilon',k'_{x}\right)\mathcal{V}_{\tau,\tau'}\left(\varepsilon,k_{x};\varepsilon',k'_{x}\right)F_{\tau'}\left(\varepsilon',k'_{x}\right)
\end{equation}
With this convention, the overall sign structure becomes
\begin{equation}\label{eq:Delta+(kx)=Delta-(-kx)}
	\Delta_{+}\left(\varepsilon;k_{x}\right)=\Delta_{-}\left(\varepsilon;-k_{x}\right)
\end{equation}
and the pairing Hamiltonian \eqref{eq:1dHamiltonian} can be written in terms of $\Delta_e$ only. The corresponding gap equation takes the same form as \eqref{eq:Deltaodd}, with $\Delta_o\to\Delta_e$.
Our theory does not distinguish energetically the sign structure on different arcs and focuses instead on the node formation. We write the nonvanishing combination appearing in the resulting Hamiltonian (7) in the main text as $\Delta=\Delta_{e/o}$ for lightness of notation.

We can diagonalize the Hamiltonian in terms of Bogoliubov quasiparticles at fixed energy $\varepsilon$  (label omitted)
\begin{equation}\label{eq:Bogoliubov}
	\left(\begin{array}{c}
		\gamma_{+,k_{x}}\\
		-\gamma_{-,-k_{x}}^{\dagger}
	\end{array}\right)=\left(\begin{array}{cc}
		u_{k_{x}}^{*} & v_{k_{x}}^{*}\\
		-v_{k_{x}} & u_{k_{x}}
	\end{array}\right)\left(\begin{array}{c}
		c_{+,k_{x}}\\
		-c_{-,-k_{x}}^{\dagger}
	\end{array}\right) \;,
\end{equation}
where the electron and hole weights $u_{k_{x}}$ and $v_{k_{x}}$ satisfy the relations $\left|u_{k_{x}}\right|^{2}+\left|v_{k_{x}}\right|^{2}=1$ and $2E\left(k_{x}\right)u_{k_{x}}^{*}v_{k_{x}}^{*}=\Delta\left(k_{x}\right)$, with the energy $E\left(\varepsilon,k_{x}\right)=\sqrt{\xi_{\varepsilon}^{2}+\left|\Delta_{}\left(\varepsilon,k_{x}\right)\right|^{2}}$. Choosing real $u$, one writes them as $u_{\varepsilon,k_{x}}=\sqrt{\frac{1}{2}\left(1+\frac{\xi_{\varepsilon}}{E\left(k_{x}\right)}\right)}$, $v_{\varepsilon,k_{x}}=e^{-i\varphi\left(k_{x}\right)}\sqrt{\frac{1}{2}\left(1-\frac{\xi_{\varepsilon}}{E\left(k_{x}\right)}\right)}$, with the order parameter phase $\varphi$ explicitly factorized. 
In terms of quasiparticles (7) in the main text takes the form 
\begin{equation}
	H=\sum_{\tau}\intop d\varepsilon\intop\frac{dk_{x}}{2\pi}n_{}\left(\varepsilon,k_{x}\right)E\left(\varepsilon,k_{x}\right)\left(\gamma_{\tau,k_{x}}^{\dagger}\gamma_{\tau,k_{x}}-\frac{1}{2}\right)
\end{equation}
With \eqref{eq:Bogoliubov}, we can evaluate the anomalous expectation value
\begin{equation}
	F_{\tau}\left(k_{x}\right)=-\tau\frac{\Delta\left(\tau k_{x}\right)}{2E\left(\tau k_{x}\right)}\tanh\frac{\beta E\left(\tau k_{x}\right)}{2}
\end{equation}
The antisymmetry \eqref{eq:antisymmF} is automatically present. Substituting into \eqref{eq:Deltaodd}, we obtain the gap equation (8) in the main text.

\subsection{On the Fermi surface}\label{sec:Around-FS}
Writing a theory purely on the Fermi surface is extremely convenient both for efficient numerical solution of the gap equations and for classifying and characterizing the allowed scattering events. At fixed energy, $\varepsilon$, an electron is identified by an arc label $\tau=\pm$ denoting the sign of $k_y$ and by its momentum along $k_x$. Taking into account momentum conservation along the surface, two kinds of processes are allowed in our model
\begin{itemize}
	\item $\tau=\tau'$ (intra-arc). The virtual phonon has a relatively small momentum
	\begin{equation}
		\boldsymbol{q}  =  \boldsymbol{k'}-\boldsymbol{k}\,=\,\left\{ k'_{x}-k_{x},\tau \left[k_y\left(\varepsilon;k'_x\right)-k_y\left(\varepsilon;k_x\right)\right]
		\right\}, 
	\end{equation}
	with $k_y$ given in \eqref{eq:ky}. This lands each electron on the same arc it is starting from.
	\item $\tau=-\tau'$ (inter-arc). These processes typically involve a large exchanged momentum
	\begin{equation}
		\boldsymbol{q}  =  \left\{k'_{x}-k_{x},-\tau \left[k_y\left(\varepsilon;k'_x\right)+k_y\left(\varepsilon;k_x\right)\right]\right\}
	\end{equation}
	of order $k_W$. In practice, when considering the separation and the extension of the arcs in 
	$\mbox{Pt}\mbox{Bi}_2$ \cite{Vocaturo2024}, these are almost vertical transitions, with the virtual phonon mainly carrying momentum along the $y$ direction.
\end{itemize}	

As scattering at the Fermi energy dominates at low temperature, we project the gap equation (8) in the main text onto the Fermi arcs at fixed energy. We neglect the energy dependence of the arc ends, of the interaction potential and of the gap function. The resulting gap equation is
\begin{equation}\label{eq:Gapeqn}
\Delta\left(k_{x}\right)=-\Lambda\intop\frac{dk'_{x}}{2\pi}\,n\left(k'_{x}\right)
V_{+,+}\left(k_{x},k'_{x}\right)\frac{\Delta_{}\left(k'_{x}\right)}{E\left(k'_{x}\right)}\tanh\frac{\beta E\left(k'_{x}\right)}{2}+\Lambda
\intop\frac{dk'_{x}}{2\pi}\,n\left(k'_{x}\right) V_{+,-}\left(k_{x},k'_x\right)\frac{\Delta_{}\left(-k'_{x}\right)}{E\left(-k'_{x}\right)}\tanh\frac{\beta E\left(-k'_{x}\right)}{2}\,,
\end{equation}
in which the energy argument is fixed and has been omitted and \mbox{$V_{\tau,\tau'}\left(k_{x},k'_{x}\right)=\mathcal{V}_{\tau,\tau'}\left(\varepsilon_F,k_{x};\varepsilon_F,k'_{x}\right)$}.
Here $\Lambda\approx \hbar\omega_D$ substitutes the omitted integration variable with the half integration range. It is absorbed in practice into the potential strength.
Noticeably, the interaction potential only depends from the absolute value of the exchanged momentum, which allows to conclude that a basis solutions of \eqref{eq:Gapeqn} has a definite parity eigenvalue.
When only intra-arc processes are retained, $V_{+,+}\ne0$, $V_{+,-}=0$, the gap equation can be linearized near the critical temperature and written as a Fredholm equation of the second kind
\begin{equation}\label{eq:gaplinintra}
	\Delta_{intra}\left(k_{x}\right)  \approx -2\intop_{-k_{end}}^{k_{end}}\frac{dk'_{x}}{2\pi}{V}_{+,+} 	\left(k_{x},k'_{x}\right)n_{}\left(k'_{x}\right) 	\Delta_{intra}\left(k'_{x}\right)\ln\left[\frac{2e^{\gamma}}{\pi}\beta\hbar\omega_{D}\right] \;,
\end{equation}
The one-dimensional kernel allows for an efficient discretization and numerical determination of its eigenvalues, see discussion in the main text. Importantly, the discretized \eqref{eq:gaplinintra} is a matrix eigenvalue equation. For a purely attractive ${V}_{+,+}$, the Perron-Frobenius theorem \cite{Meyer,Jentzsch1912} ensures that a non-degenerate eigenvalue of largest magnitude exists and that the corresponding eigenvector has positive elements. The largest gap is therefore even under reflection $k_x\to-k_x$. In particular, for an eigenvector $\phi_j$ with eigenvalue $v_{intra}^{(j)}$, it must be true that
\begin{equation}\label{eq:egvalintra}
	1=-2 v_{intra}^{(j)} \ln\left[\frac{2e^{\gamma}}{\pi}\beta\hbar\omega_{D}\right] \;.
\end{equation}
The highest critical temperature corresponds to (9) in the main text. 
We now consider \eqref{eq:Gapeqn} when only inter-arc processes $V_{+,+}=0$, $V_{+,-}\ne0$ are included. If we retain vertical transitions only $k'_x=k_x$, \eqref{eq:Gapeqn} reduces to an algebraic relation which only admits odd solutions. More generally, the gap equation is linearized around the critical temperature as
\begin{equation}\label{eq:gaplininter}
	\Delta_{inter}\left(k_{x}\right)  \approx 2\varpi\intop_{-k_{end}}^{k_{end}}\frac{dk'_{x}}{2\pi}{V}_{+,-} \left(k_{x},k'_{x}\right)n\left(k'_{x}\right) 	\Delta_{inter}\left(k'_{x}\right)\ln\left[\frac{2e^{\gamma}}{\pi}\beta\hbar\omega_{D}\right] \;,
\end{equation}
where $\varpi=\pm1$ is the eigenvalue under inversion along the arc of the gap solution
$\Delta_{inter}\left(-k_x\right)=\varpi \Delta_{inter}\left(k_x\right)$.
The eigenvalue condition becomes
\begin{equation}\label{eq:egvalinter}
	1=2 \varpi v_{inter}^{(j)} \ln\left[\frac{2e^{\gamma}}{\pi}\beta\hbar\omega_{D}\right] \;.
\end{equation}


\bibliography{WSS}

\end{document}